\documentclass[useAMS,usegraphicx,usenatbib]{mn2e}

\usepackage{amssymb,amsmath}
\usepackage[normalem]{ulem}

\usepackage{natbib}

\newcommand{\Bf}{{magnetic field}}

\newcommand\eg{\textit{e.g.}}

\newcommand\ie{\textit{i.e.}}

\newcommand{\be}{\begin{equation}} 
\newcommand{\ee}{\end{equation}}

\newcommand{\ba}{\begin{eqnarray}}
\newcommand{\ea}{\end{eqnarray}}

\title[Mass-loading of bow shock PWN]{Mass-loading of bow shock pulsar wind nebulae}
\author[G. Morlino, M. Lyutikov and M.~J. Vorster]{
	G. Morlino,$^{1,2}$\thanks{E-mail: giovanni.morlino@gssi.infn.it}
	M. Lyutikov,$^1$\thanks{E-mail: lyutikov@purdue.edu} and
	M. Vorster$^{1}$\thanks{E-mail: mvorster@purdue.edu}	
\\
	{$^1$ \it Department of Physics, Purdue University, 525 Northwestern Avenue, West Lafayette, IN 47907-2036, USA.}\\
	{$^2$ \it INFN -- Gran Sasso Science Institute, viale F. Crispi 7, 67100 L'Aquila, Italy.}
}

\begin{document}

\date{Accepted 2015 September 21.  Received 2015 September 21; in original form 2015 May 6.}


\maketitle

\label{firstpage}

\begin{abstract}
We investigate  the dynamics of bow shock nebulae created by pulsars moving supersonically through a partially ionized interstellar medium.  A fraction of interstellar neutral hydrogen atoms penetrating into the tail region of a pulsar wind will undergo photo-ionization due to the UV light emitted by the nebula, with the resulting mass loading dramatically changing the flow dynamics of the light leptonic pulsar wind.  Using a quasi 1-D hydrodynamic model of both  non-relativistic and relativistic flow, and focusing on scales much larger than the stand-off distance, we find that if a relatively small density of neutral hydrogen, as low as $10^{-4}\,\text{cm}^{-3}$, penetrate inside the pulsar wind, this is sufficient to strongly affect the tail flow.  Mass loading leads to the fast expansion of the pulsar wind tail, making the tail flow intrinsically non-stationary.  The shapes predicted for the bow shock nebulae compare well with observations, both in H$\alpha$ and X-rays.  
\end{abstract}


\section{Introduction}
 \label{sec:intro}
 
It has been estimated that between $10\%$ and $50\%$ of pulsars are born with kick velocities $V_{\rm NS} \gtrsim 500\,\text{km}\,\text{s}^{-1}$ \citep{Cordes98,Arzoumanian2002}. These pulsars will escape from their associated supernova remnants into the cooler, external  interstellar medium (ISM) in less than $20\,\text{kyr}$ \citep{Arzoumanian2002}.  As this time scale is sufficiently short, the pulsars are still capable of producing  powerful relativistic winds.  Furthermore, comparison with typical sound speeds in the ISM, $c_{s, \rm ISM} = 10-100 \,\text{km}\,\text{s}^{-1}$, shows that the pulsars are moving with highly supersonic velocities.  The interaction of the pulsar's wind with the ISM produces a bow shock nebula with an extended tail. If a pulsar is moving through a partially ionized medium, the bow shock nebula can be detected by the characteristic H$\alpha$ emission resulting from the collisional  and/or charge-exchange  excitation of neutral hydrogen atoms in the post-shock flows and the subsequent emission via bound-bound transitions \citep{Chevalier80}.  To date, nine such bow shock nebulae have been discovered, including three around $\gamma$-ray pulsars \citep{Brownsberger14}. 

\begin{figure}
\begin{center}
\includegraphics[width=0.47\textwidth]{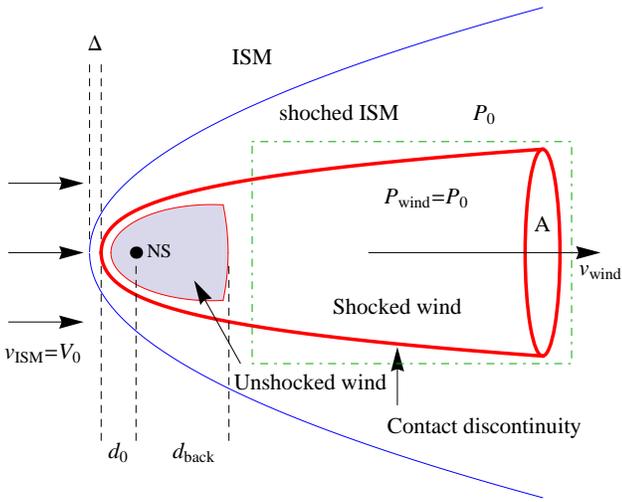}
\end{center}
\caption{Schematic illustration of a bow shock nebula propagating through {\it fully ionized} ISM, as seen in the rest frame of the pulsar. The dot-dashed rectangle shows the region zoomed in Fig.\ref{fig:sketch2}.}
\label{fig:sketch1}
\end{figure}

Hydrodynamic (and hydromagnetic) models \citep[\eg][]{Bucciantini02a} of bow shock nebulae predict the formation of a smooth two shock structure schematically shown in Fig. \ref{fig:sketch1}: a forward shock in the ISM separated by a contact discontinuity from a termination shock in the pulsar wind.  In the head of the nebula the shock and the contact discontinuity are situated at a distance given by Eq. (\ref{eq:d0}), corresponding to the position where the ram pressure of the ISM balances the pulsar wind pressure.  The flow structure in the head of the nebula is  reasonably well understood, especially in the limit of strong shocks (\ie, when the pulsar velocity is much larger than the ISM sound speed) and neglecting the internal structure of the shocked layer \citep{Wilkin96}\footnote{This model is very realistic when the system cools efficiently, otherwise the pressure of the shocked ISM needs to be taken into account.}.

These models further predict that the pulsar wind terminates at a Mach disk located approximately at a distance $d_{\rm back} = M_{\rm NS} d_0$ behind the pulsar (where $M_{\rm NS} = V_{\rm NS}/c_{s, \rm ISM}$  is the Mach number of the pulsar moving through the ISM). At approximately the same distance the oblique forward shock turns into a Mach cone with an opening angle $\sim 1/M_{\rm NS}$. For distances larger than $d_{\rm back}$ the flow in the tail of the nebula is smooth and nearly cylindrical, although some models predict the development of shear flow instabilities (\eg, Kelvin-Helmholtz instabilities).  

\begin{figure}
\begin{center}
\includegraphics[width=0.47\textwidth]{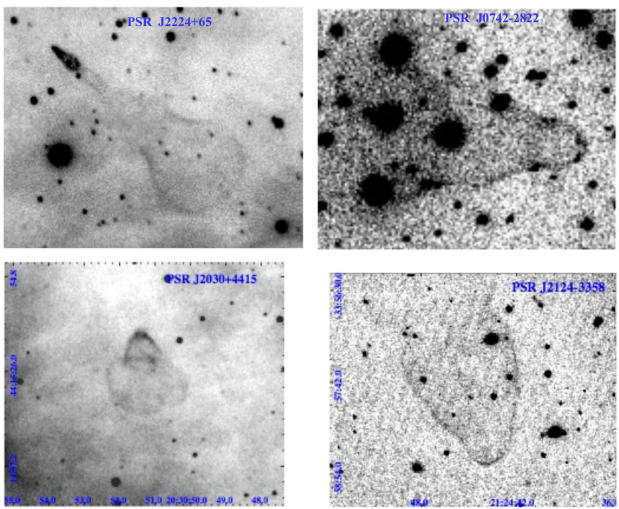}
\end{center}
\caption{Montage of H$\alpha$ images of optical bow shocks associated with pulsar wind nebulae. Shown are J2224+65, the so-called {\it Guitar nebula} \citep{Chatterjee02}, J0742-2822, J2030+4415, and J2124-3348 \citep{Brownsberger14}.}
\label{fig:Halpha_PWNe}
\end{figure}

In contrast to these numerical models, H$\alpha$, radio and X-rays observations show that the morphologies of bow shock nebulae are significantly more complicated.  More specifically, observations reveal that the tails of bow shock nebulae have a highly irregular morphology, with Fig.\ref{fig:Halpha_PWNe} showing four such examples.  All these nebulae have a characteristic \emph{``head-and-shoulder'' structure}, with the smooth bow shock in the head not evolving into a quasi-conical or  quasi-cylindrical shape, but instead showing a sudden sideways expansion(s).  Arguably the most famous example is the {\it Guitar nebula} powered by the pulsar PSR B2224+65 (top left panel in Fig.\ref{fig:Halpha_PWNe}).  As the name suggests, this nebula has a guitar-like shape with a bright head, a faint neck, and a body consisting of several larger bubbles.   

Although the morphologies of these nebulae vary from source to source, there are a number of common features which, in our view, not only reflect \emph{the intrinsic dynamical  properties of the flows}, but which are also independent of the subtle details of both the pulsar winds (\eg, the relative orientation of the velocity and spin axis) and of the ISM. 
We stress  the fact that {\it all} bow shock nebulae show  {\it qualitatively similar morphological features} not expected from simple fluid models.  In the X-ray and radio bands the tails show highly non-trivial morphologies with quasi-periodic variations in the intensity \citep[\eg][]{Kargaltsev08}.  For example, in the case of the  Guitar nebula the tail shows {\it quasi-periodic} bubble-like structures \citep{Kerkwijk08}.  

These peculiar tail shapes have been interpreted as the result of density variations in the ISM \citep{Romani97,Vigelius07}.  However, based on the following considerations, we find this explanation unsatisfactory: ({\it i}) all tails show {\it similar} morphological variations (see Fig.\ref{fig:Halpha_PWNe}); ({\it ii}) a common characteristic of these bow shock nebulae is that they are all highly symmetric with respect to the direction of motion of the pulsar - this is not expected if variations are due to the external medium; ({\it iii}) morphological features in H$\alpha$, radio and X-rays are quasi-periodic -  this is also not expected from random ISM density variations. 
From these observations we conclude that the {\it  peculiar morphological features result from the internal dynamics of the pulsar wind}, rather than through inhomogeneities in the ISM.  

\cite{Kerkwijk08} have also previously proposed that the morphology of the Guitar nebula could be explained by (unidentified) instabilities in the jet-like flow of pulsar material away from the bow shock. 
Alternatively, \citet{Bucciantini01} and \citet{Bucciantini02b} have suggested that the mass loading of pulsar wind nebulae may strongly affect their dynamics. These authors have shown that a non-negligible fraction of neutral atoms can cross the shocked ISM behind the bow shock without undergoing any interaction, thereby enabling these atoms to propagate into the pulsar wind region.  Once inside the wind, neutral hydrogen can be ionized by UV or X photons emitted by the nebula, and possibly  by collisions with relativistic electrons and positrons, resulting in a net mass loading of the wind.  

In order to study this scenario, \citet{Bucciantini01} and \citet{Bucciantini02b} extended the thin-layer approximation used to model cometary nebulae \citep{Bandiera93, Wilkin96, Wilkin00}.  The thin-layer approximation is conceptually analogous to a 1-D model as it neglects the thickness of the nebula, while all quantities depend only on the distance from the apex.  Despite the above-mentioned simplifications, these models provide a good description of the head region of the nebulae in terms of shape, hydrogen penetration length scale, and H$\alpha$ luminosity, as was later confirmed by more accurate 2-D axisymmetric simulations, both in the hydrodynamic (HD) regime \citep{Bucciantini02a, Gaensler04} and in the relativistic magnetohydrodynamic (MHD) regime \citep{Bucciantini05}. Using a 3-D model, \citet{Vigelius07} was able to extend the study of these systems by also taking into account either a non-uniform ambient medium, or the anisotropy of the pulsar wind energy flux. However, none of these models are able to explain the peculiar morphology of the H$\alpha$ emission often observed in the tail regions of bow shock nebulae.

While the above-mentioned studies focused primarily on the head of bow shock nebulae, the aim of the present paper is {\it to investigate the effect of neutral hydrogen on the tail region of these nebulae}. The question we would like to investigate is whether the mass loading of neutral hydrogen in the pulsar wind can explain the peculiar morphology observed at H$\alpha$, radio and X-ray energies. In order to focus on the effect of mass loading on the evolution of bow shock nebulae, complications introduced by magnetic field pressure (and topology) are neglected in the present paper.  These aspects are indeed necessary for a comprehensive and realistic treatment of the problem, and will be the subject of a future study.   

At this point the question arises as to whether one can use observations of the heliopshere to understand the problem formulated in the previous paragraph.  Although mass loading plays an important role in the dynamics of the solar wind \citep{bkk71,bara90,zank99}, there are a number of key differences between this scenario and the pulsar wind scenario.  Firstly, the velocity of the Sun through the ISM is, most likely, weakly sub-fast magnetosonic \citep{2012Sci...336.1291M}, whereas the pulsar's motion is highly supersonic; secondly, the pulsar wind is very light - composed of lepton pairs - and one would therefore expect mass loading to have a greater effect; thirdly, most of the research related to mass loading in the solar wind concentrated on flow in the head region, while we are interested in the large scale dynamics of the tail flows.

The outline of the rest of the paper is as follows: in \S\ref{sec:loading} we discuss how neutral hydrogen penetrates into the pulsar wind, summarising the possible interaction processes (charge-exchange, photo-ionization and collisions). In particular, we show that the ionization of neutrals inside the wind is mainly  due to UV photons emitted by the nebula.  We stress here that in the whole paper with ``neutrals'' we will always refer to neutral hydrogen, even if, in principle, neutral Helium can play a similar role. We will also briefly argue at the end of \S~\ref{sec:loading} why ions from the ISM should play no role in mass loading.
The analytical, non-relativistic HD model used for the study is presented in \S\ref{sec:non-rel-model} and \S\ref{sec:limits-NR}, while in \S\ref{sec:rel-model} we develop a similar model but for a relativistic wind. In \S\ref{sec:fit} we provide a simple visual model fit to the bow shock nebula PSR J0742-2822, while we qualitatively discuss in \S\ref{sec:magnetic} how the presence of a magnetic field inside the wind can modify the wind dynamics. Lastly, a summary of the main results can be found in \S\ref{sec:conc}. 

\section{Structure and  relevant scales of a PWN\lowercase{e} confined by a partially ionized medium} 
 \label{sec:loading}

Fig.(\ref{fig:sketch1}) shows the typical structure produced by a pulsar (or a normal star) moving supersonically through the  ISM, as seen in the rest frame of the pulsar, when the effect of neutrals is not taken into account. 
Such bow shock structures are preferentially produced when the pulsar propagates through the \emph{warm phase} of the ISM, whose typical temperature is $T\sim 6\cdot10^3$-$10^4$ K. In this case the sound speed is $c_s\sim 10$ km/s, hence the pulsar's Mach number is $\gg 1$. Conversely, the \emph{hot phase} of the ISM has $T\sim 10^6$ K and $c_s\sim 100$ km/s, implying a Mach number close to one. In the hot phase the ISM is totally ionised, while in the worm one the presence of neutral Hydrogen cannot be neglected. In fact the typical density of the warm ISM is 0.2-0.5 cm$^{-3}$ and the ionized fraction is estimated to range between 0.007-0.05 for the warm neutral medium (WNM), and 0.6-0.9 for the ionised neutral medium (WIM) \cite[see, e.g.,][Table 1]{Jean09}.

The distance, $d_0$, between the pulsar and the contact discontinuity (CD) (formed between the shocked ISM and the shocked wind) is obtained by equating the wind pressure with the bulk pressure of the ISM:
\begin{equation} \label{eq:d0}
 d_0= \left( \frac{\mathcal{L}_{w}}{4 \pi V_{\rm NS}^2  \rho_{\rm ISM} c}  \right)^{\frac{1}{2} }
       = 1.3 \times 10^{16} \mathcal{L}_{w,34}^{1/2} \, V_{300}^{-1} \, n_{\rm ISM,-1}^{-1/2} \, {\rm cm} \,,
\end{equation}
where $\mathcal{L}_{w}= 10^{34} \mathcal{L}_{w,34}\,\text{erg}\,\text{s}^{-1}$ is the pulsar luminosity, $V_{\rm NS}= 300\, V_{300}\,\text{km}\,\text{s}^{-1}$ is its peculiar velocity, and $\rho_{\rm ISM}=m_p n_{\rm ISM}$ is the density of the dragged component of the ambient medium, expressed in units of $n_{\rm ISM}= 0.1\, n_{\rm ISM,-1}\,\text{cm}^{-3}$. 

In order for mass loading to play a role in the dynamic evolution of bow shock nebulae, neutrals are required to cross $\Delta$ (the distance between the bow shock and the contact discontinuity) and penetrate into the wind region. Hence the interaction length inside the shocked ISM, $\lambda$, must be larger than $\Delta$.  \citet{Chen96} estimated that $\Delta= 5/16 \,d_0$, a value that was later confirmed by \citet{Bucciantini05} using numerical simulations (note that $\Delta$ is smaller than $d_0$). 

At this stage it is important to note that different physical processes will lead to the interaction of neutrals in the shocked ISM than in the pulsar wind. In the next section it will be shown that a significant amount of neutrals can undergo charge exchange (CE) with protons in the shocked ISM and collisional ionization with electrons (specifically in the head of the bow shock system), while the collisional ionization of neutrals by protons can be neglected when $V_{\rm NS} \lesssim 300\,\text{km}\,\text{s}^{-1}$.  Despite the neutrals undergoing CE and ionization, it will further be shown that a dynamically important fraction of neutrals can still penetrate into the pulsar wind in their original state.

The pulsar wind most likely consists of only electrons and positrons \citep{1975ApJ...196...51R}, and CE is therefore not possible. Rather, ionization of neutrals can only occur through photo-ionization or through the collision with relativistic electrons and positrons. The former process is discussed in \S\ref{sec:ph-ionization} where it is shown that a significant amount of neutrals can be photo-ionized through the non-thermal emission emitted by the pulsar wind, while in \S\ref{sec:collisions} we show that collisional ionization can be neglected. As photo-ionization is important for the pulsar wind, one may ask whether this process is also important for the shocked ISM in the head of the nebula. Section \S\ref{sec:ph-ionization-ISM} will therefore be used to discuss when photo-ionization in the shocked ISM can be neglected. 
Lastly, for the sake of completeness, we will discuss in \S\ref{sec:rel_wind} why the ionization of neutrals inside the unshocked pulsar wind can be neglected.

One may wonder whether ions from the ISM, rather than neutrals, can penetrate inside the wind directly, resulting in a net mass loading of the wind. The main issue of this hypothesis is that ions are attached to magnetic field lines which does not cross the contact discontinuity. In principle ions could diffuse perpendicular to the field lines and enter inside the wind. Nevertheless a simple estimate excludes this possibility: using the Bohm diffusion coefficient, $D_B = v r_L/3$, as an upper limit for the perpendicular diffusion, we can estimate the time needed for a thermal proton to cross the typical distance $d_0$, which is given by $t_{\rm cross}= d_0^2/D_B$. Using $B= 1 \mu$G, $d_0= 10^{16}$ cm and $v= 100$ km s$^{-1}$, we get  $t_{\rm cross}\approx10^9$ yr. Hence the diffusion of ions is too slow and can be neglected.
An alternative way for ions to penetrate inside the wind is through  shear flow instabilities that can develop at the contact discontinuity. Nevertheless, the effects of instabilities will not be limited to the simple injection of ions, but will mix the ISM materials with the pulsar wind resulting in a complex tale structure and probably to its disruption. In our knowledge this scenario has never been studied in details, and it is far beyond the aim of the present paper.

\begin{table*}
\caption{\label{tab:2} Tabulated quantities, in order of appearance: pulsar name, spin-down power, non-thermal X-ray emission from the pulsar, non-thermal X-ray flux from nebula, photon index of nebula X-ray emission, ISM density.  References: (1) \citet{Brownsberger14}, (2) \citet{Abdo2013}, (3) \citet{Hui2007}, (4) \citet{Romani2010}, (5) \citet{Stappers2003}, (6) \citet{Hui2006}.  The values of $L_{\rm{X}}$ marked with an $^*$ are calculated using the relation $\text{log}_{10}L_{\rm{X,pwn}}=1.51\,\text{log}_{10}\dot{E} - 21.4$ derived by \citet{Kargaltsev2012} from \emph{Chandra} observations.   }
\begin{center}
\begin{tabular}{lcccccc}
\hline
\hline
Pulsar & $\dot{E}_{34}$ & $ L_{\rm{X,pul}}$ & $ L_{\rm{X,pwn}}$ & $\Gamma_{\rm{X,pwn}}$ & $n_{\rm{ISM}}$ & Reference\\
       & [$\text{erg}\,\text{s}^{-1}$] & [$\text{erg}\,\text{s}^{-1}$] & [$\text{erg}\,\text{s}^{-1}$] & & [$\text{cm}^{-3}$] &\\
\hline
J0437--4715 & $0.55$ & $2.4\times 10^{30}$ & $3.5\times 10^{29}$ $^*$ & -- & $0.21$ & 1, 2 \\
J0742--2822 & $19.0$ & $<9.6\times 10^{30}$ & $7.4\times 10^{31}$ $^*$ & -- & $0.28$ & 1, 2 \\
J1509--5850 & $68.2$ & $2.4\times 10^{32}$ & $(1.3-2.7) \times 10^{32}$ & $1.3^{+0.8}_{-0.4}$ & $6.14$ & 1, 2, 3\\
J1741--2054 & $12.6$ & $3.5\times 10^{30}$ & $4.0\times 10^{31}$ $^*$ & $1.6\pm 0.2$ & $1.44$ &  1, 2, 4\\
J1856--3754 & $3\times 10^{-4}$ & -- & $4.2\times 10^{34}$ $^*$ & -- & $0.05$ & 1, 2 \\
J1959+2048  & $21.9$ & $5.2\times 10^{31}$ & $1.9\times 10^{31}$ & $1.5\pm 0.5$ & $0.02$ & 1, 2, 5\\
J2030+4415  & $2.90$ & $2.7\times 10^{31}$ & $4.3\times 10^{30}$ $^*$ & -- & $2.69$ & 1, 2\\
J2124--3358 & $0.68$ & $8.6\times 10^{29}$ & $10^{29}$ & $2.2\pm 0.4$ & $0.47$ & 1, 2, 6 \\
J2225+6535  & $0.16$ & -- & $5.5\times 10^{28}$ $^*$ & -- & $1.43$ & 1, 2\\
\hline
\end{tabular}
\end{center}
\end{table*}

\subsection{Interaction of neutrals in the shocked ISM} \label{sec:shocked_ISM}
After crossing the bow shock, neutral atoms can interact with the shocked protons. The interaction length is given by 
\begin{equation} \label{eq:L_coll}
 \lambda= \frac{V_{\rm NS}}{ X_{\rm ion} n_{\rm ISM} r_c \langle \sigma(v_{\rm rel}) v_{\rm rel} \rangle} \,,
\end{equation}
where $X_{\rm ion}$ is the ionization fraction of the ISM, $r_c$ is the compression ratio of the bow shock, $\sigma(v_{\rm rel})$ is the relevant cross section of the process under consideration, and $\langle \sigma v_{rel} \rangle$ is the collision rate averaged over the ion distribution function.  When the ion distribution is a Maxwellian, $\langle \sigma v_{rel} \rangle$ is well approximated (within 20\%) by the expression \citep{Zank96,Blasi12} 
\begin{equation}
\langle \sigma v_{rel} \rangle \approx \sigma(U^*) U^*,
\end{equation}
where 
\begin{equation}
U^*= \sqrt{ \frac{8}{\pi}  \frac{2 k_B T}{m_p} }
\end{equation}
is the average, relative speed between the incoming hydrogen atom and ions ($T$ is the temperature of the shocked ISM determined by the Rankine-Hugoniot jump conditions, assumed to be $\gg T_{\rm ISM}$).  Using the fiducial values $n_{\rm ISM}=0.1\,\text{cm}^{-3}$ and $V_{\rm NS}= 300\,\text{km}\,\text{s}^{-1}$, together with an ionization fraction of 90\% and $r_c=4$ (the typical value for strong shocks), leads to the following estimates for the mean free paths
\begin{align} 
 \lambda_{{\rm ion},p} & \approx 3.0 \times 10^{20} \, \rm cm   \label{eq:L_ion_p} \\
 \lambda_{{\rm ion},e} & \approx 2.2 \times 10^{16} \, \rm cm   \label{eq:L_ion_e}\\
 \lambda_{\rm CE} & \approx 1.5 \times 10^{15} \, \rm cm \,     \label{eq:L_CE}
\end{align} 
for the ionization due to collisions with protons, electrons and CE, respectively.  Note that $\lambda_{{\rm ion},e}$ has been calculated under the assumption that the electrons downstream of the shock equilibrate rapidly with protons, thereby acquiring the same temperature.  If this assumption does not hold, the collisional length scale for ionization due to electrons can become much larger than the value reported in Eq.~(\ref{eq:L_ion_e}). 
In addition, the values (\ref{eq:L_ion_p})-(\ref{eq:L_CE}) are to be taken as lower limits as they are valid just ahead of the nebula, where the compression ratio and the temperature obtain their maximum values. 

From these estimates it follows that only a negligible fraction of the neutral hydrogen will be collisionally ionized by electrons, whereas a significant fraction of neutrals will undergo CE.  The neutrals resulting from a CE event will have a bulk speed and a temperature that are close to that of the protons in the shocked ISM.  This implies that  the newly formed neutrals tend to be dragged with the shocked protons along a direction parallel to the contact discontinuity. Nevertheless, the CE process produces a diffusion of neutrals in the nose of the nebula and it may still be possible for the newly formed neutrals to enter the wind region, provided that their diffusion velocity perpendicular to the contact discontinuity is of the same order or larger than their velocity parallel to the contact discontinuity. This is a complication that will not be addressed in the present paper but is essential to estimate the correct amount of neutrals that can penetrate into the wind. 

Although a large number of neutrals will be lost due to CE, a minimum fraction of neutrals proportional to $\exp[- \Delta/(\lambda_{\rm CE}+\lambda_{\rm ion})]$ will cross the shocked ISM region without suffering any interaction and will enter the pulsar wind in their original state.  These neutrals will not influence the wind structure until they are ionized, either through collisions with relativistic electrons and positrons, or through photo-ionization with photons emitted by the nebula or by the pulsar. In \S\ref{sec:ph-ionization} we show that photo-ionization is the dominant process in pulsar wind nebulae, while we demonstrate in \S\ref{sec:collisions} that collisional ionization can be neglected.  However, we first discuss the photo-ionization ahead of the nebula and in the shocked ISM.

\subsection{Photo-ionization outside the pulsar wind} \label{sec:ph-ionization-ISM}
There are three different sources of photons to account for: thermal and non-thermal radiation from the pulsar, and non-thermal radiation from the nebula. The thermal radiation from the pulsar has been shown to be negligible \cite[see][Eq.(25) and discussion below]{Bucciantini01}. On the other hand, non-thermal emission from both the pulsar and the nebula can play a role as the non-thermal pulsar luminosity is generally comparable to the luminosity of the nebula.

When the radiation emitted by the nebula has a sufficient luminosity, incoming neutrals can be ionized before crossing the bow shock, and one would consequently not expect any effect from mass loading. Conversely, the requirement that hydrogen atoms are not fully ionized at the bow shock imposes an upper limit on the total ionizing luminosity.  We derive this limit closely following a similar derivation given by \cite{vanKK01}.  

Assuming a spherical symmetry for the emission emitted by the nebula, the ionization fraction $\xi_+$ at a distance $r$ from the center of the nebula is
\begin{equation} \label{eq:ion_frac}
 \frac{d\xi_+(r)}{dt} = (1-\xi_+) \bar \sigma_{\rm ph} \frac{N_{\rm ph} e^{-\tau(r)}}{4 \pi r^2} - \xi_+ n_e \alpha_{\rm rec} \,,
\end{equation}
where $N_{\rm ph}$ is the number of ionizing photons emitted by the nebula per unit time, and $\bar \sigma_{\rm ph}= N_{\rm ph}^{-1}\,\int_I^{\infty} \sigma_{\rm ph}(\nu) n_{ph}(\nu) d\nu$ is the photo-ionization cross section averaged over the photon distribution. The photo-ionization cross section is 
\begin{equation} \label{eq:sigma_ph}
  \sigma_{\rm ph}(\nu)= 64 \alpha_{\rm fin}^{-3} \sigma_T \left(I / h\nu \right)^{7/2}
  				          = 10^{-16} (h\nu/{\rm Ryd})^{-7/2} \, \rm cm^{2},
\end{equation} 
with $\sigma_T$ the Thompson cross section and $I=1$ Ryd the ionization potential. As $\sigma_{\rm ph}$ decreases rapidly with photon energy, the only relevant photons are those with energy $\gtrsim I$.
Solving Eq.(\ref{eq:ion_frac}) in the rest frame of the pulsar where the plasma is moving in the positive direction along the $x$-axis with velocity $V_{\rm NS}$, one may write $dx=V_{\rm NS} dt$. If the extinction and recombination of protons are neglected, Eq.(\ref{eq:ion_frac}), written in cylindrical coordinates $(x,\rho)$, simplifies to
\begin{equation} \label{eq:ion_frac_cyl}
 \frac{d\xi_+}{1-\xi_+} =  \frac{\bar\sigma_{\rm ph} N_{\rm ph}}{4 \pi V_{\rm NS}} \frac{dx}{\rho^2+x^2} \,.
\end{equation}
The solution  is straightforward and reads
\begin{equation} \label{eq:ion_frac_sol}
 \xi_+(x) =  1- (1-\xi_{+,0}) \exp \left\{ -\frac{r_0}{\rho} 
                  \left[\frac{\pi}{2} - \arctan \left(\frac{x}{\rho}\right) \right]  \right\} \,,
\end{equation}
where $\xi_{+,0}$ is the original ionization fraction of the ISM far from the nebula and $r_0\equiv \bar \sigma_{\rm ph} N_{\rm ph}/(4 \pi V_{\rm NS})$ defines the typical distance where ionization is effective. 
As pointed out by \cite{vanKK01}, Eq.(\ref{eq:ion_frac_sol}) describes an ionization fraction which, for fixed impact parameter $\rho$, smoothly increases as one approaches the nebula from the left in Fig.(\ref{fig:sketch1}), strongly increases for $\rho\sim r_0 $, before approaching the asymptotic value $1- (1-\xi_{+,0}) e^{-\pi r_0/\rho}$ as one moves away from the pulsar towards the right.

In order to estimate the value of $r_0$, it is useful to express the product $\bar \sigma_{\rm ph} \cdot N_{\rm ph}$ using the X-ray luminosity of the nebula, $\mathcal{L}_X$, which we define as the luminosity in the Chandra energy range, \emph{i.e.}, between $\epsilon_1=0.5$ keV and $\epsilon_2=8$ keV. 
Assuming that the non-thermal emission from the nebula ranging between UV and X-ray energies is a single power law with a photon index $\Gamma$, the value of $r_0$ can be written as
\begin{equation} \label{eq:r0}
 r_0  = \frac{\bar \sigma_{\rm ph} N_{\rm ph}}{4 \pi V_{\rm NS}} 
 	= 1.33 \cdot 10^{16} \mathcal{F}(\Gamma) \mathcal{L}_{X,30} V_{300}^{-1} \, \rm cm \,,
\end{equation} 
where $\mathcal{L}_{X,30}$ is the X-ray luminosity in units of $10^{30}\,\text{erg}\,\text{s}^{-1}$ and $V_{300}$ is the pulsar speed in units of $300\,\text{km}\,\text{s}^{-1}$, while $\mathcal{F}$ is a dimensionless function that depends only on the photon index
\begin{equation} \label{eq:F_Gamma}
  \mathcal{F}(\Gamma)= I^{9/2} \frac{2\Gamma-4}{2\Gamma+5} \,
  		\frac{[\epsilon^{-5/2-\Gamma}]_{I}^{\epsilon_2}}{[\epsilon^{2-\Gamma}]_{\epsilon_1}^{\epsilon_2}}  \,.
\end{equation} 
The typical values of $\Gamma$ inferred for PWNe detected in X-rays are $\Gamma\geq 1.5$\footnote{It should be kept in mind that a value of $\Gamma\sim 1.5$ is expected if the particles responsible for the non-thermal X-rays are produced through Fermi acceleration.} \citep[see also, \eg][]{Kargaltsev08}. It follows that $\mathcal{F} \geq 2\times 10^{-3}$, with $\mathcal{F}$ reaching unity when $\Gamma=2.55$. Eq.(\ref{eq:r0}) shows that $r_0$ can easily be smaller than the stand-off distance, $d_0$, implying that the majority of neutrals reach the bow shock without having been ionized.
It is useful to express the condition $r_0<d_0$ in terms of an upper limit for the X-ray efficiency of the nebula, defined as $\eta_X\equiv \mathcal{L}_X/\mathcal{L}_w$:
\begin{equation} \label{eq:L_max}
 r_0<d_0  \; \Rightarrow \;
 \eta_X < 10^{-4} \, {\mathcal{F}(\Gamma)}^{-1} \, \mathcal{L}_{w,34}^{-1/2} \, n_{\rm ISM,-1}^{-1/2} \,.
\end{equation}
Typical measured values for PWNe are $\eta_X\approx 10^{-3}$, although large deviations from this value are observed. 
For H$\alpha$ emitting bow shock nebulae it is known that hydrogen atoms are present at the position of the bow shock, therefore the inequality (\ref{eq:L_max}) has to be satisfied. In fact, for PWNe emitting both H$\alpha$ bow shock and X-ray radiation, the estimated value of $\eta_X$ is closer to $10^{-4}$ (see values reported in Table~\ref{tab:1}), and the inequality (\ref{eq:L_max}) is thus satisfied.

\subsection{Photo-ionization inside the pulsar wind} \label{sec:ph-ionization}
The photo-ionization of atoms inside the wind determines the rate of mass loading. In order to estimate this rate, the value of the photon density inside the nebula is required.  For simplicity it is assumed that the wind is a cylinder with a cross section $\pi d_0^2$ and a length $R$, emitting radiation uniformly.  The photon number density inside the wind is then given by
\begin{equation} \label{eq:n_ph}
 n_{\rm ph,w} = \frac{N_{\rm ph} \langle l \rangle}{\pi d_0^2 R c} \,.
\end{equation}
The value of $R$ can vary greatly from one bow shock nebula to another, but in general $R\gg d_0$. The quantity $\langle l \rangle$ is the mean length traversed by a photon before escaping the nebula. When the presence of neutrals is neglected, PWNe are transparent to UV radiation because the $e^+/e^-$ pair density is very small and the Compton scattering mean free path is greater than the size of the nebula\footnote{This conclusion could be different only for very young PWNe which have a larger $e^\pm$ density \cite[see][for details]{Atoyan96}.}. 
Hence for geometrical reasons we can assume $\langle l \rangle \sim d_0$. If we account for the presence of neutrals inside the wind, the mean free path of photons due to ionization is $l_{\rm mfp} = 1/(\sigma_{\rm ph} n_{N})$, where $\sigma_{\rm ph}$ is given by Eq.(\ref{eq:sigma_ph}) and $n_{N}$ is the neutral density inside the wind. We do expect $n_{N} < n_{N,\rm ISM} \sim 0.1$ cm$^{-3}$, hence $l_{\rm mfp} > 10^{17}$ cm. In conclusion, also when we account for the presence of neutrals, the assumption $\langle l \rangle \sim d_0$ remains a good estimate and the ratio $R/\langle l \rangle$ is  $\gg 1$. Using Eq.~(\ref{eq:n_ph}), the ionization length of neutrals inside the wind is estimated to be
\begin{eqnarray} \label{eq:L_ph_ion}
 \lambda_{\rm ph} =  \frac{V_{\rm NS}}{n_{\rm ph,w} \, \bar\sigma_{\rm ph} c}  \hspace{5cm} \nonumber \\
    \hspace{0.5cm} = 3.2 \cdot 10^{15} \, \frac{R}{\langle l \rangle} 
		                  \left(\frac{\eta_X}{10^{-4}} \right)^{-1} V_{300}^{-1} \,
		                  n_{\rm ISM,-1}^{-1} \, \mathcal{F}(\Gamma)^{-1}\, \rm cm \,,
\end{eqnarray}
where the second equality has been obtained using Eqs.~(\ref{eq:d0}) and (\ref{eq:r0}).
One important comment is in order: the ionization length scale, $\lambda_{\rm ph}$, is ultimately not the length scale that determines whether neutrals will influence the dynamics of the wind. $\lambda_{\rm ph}$ is an estimate of the length scale required to ionize the largest fraction of neutrals.  The quantity more pertinent to the  dynamics of the wind is the relative change in the density of the wind induced by mass loading.  Since the pulsar wind is light,  consisting of electron-positron pairs,  only a small fraction of neutrals are required to be ionized in order for these neutrals to become dynamically important.  The more important parameter is therefore  the length scale where the dynamics of the wind is dominated by the ionized protons rather than by the electron-positron pairs.  
Thus, we define $\lambda_{\rm ML,ph}$ as the length scale where the loaded mass is equal to the initial mass of the wind, \emph{i.e.}, 
\begin{equation} \label{eq:Lmass_load2}
 \lambda_{\rm ML,ph} 
   =  \frac{\rho_e}{\rho_N} \frac{V_{\rm wind}}{n_{\rm ph,w}  \bar\sigma_{\rm ph}\, c } 
   =  \frac{n_e m_e}{n_N m_p} \frac{V_{\rm wind}}{V_{\rm NS}} \lambda_{\rm ph}  \,.
 \end{equation}
A numerical estimate of $\lambda_{\rm ML,ph}$  can be found using Eqs.(\ref{eq:L_ph_ion}) to evaluate $\lambda_{\rm ph}$, and estimating the electron density of the wind, $n_e$, by equating the luminosity of the pulsar with the energy flux in the shocked wind, \emph{i.e.}, $\dot{M} \equiv \mathcal{L}_{w}/(\gamma c^2) = n_e m_e \pi d_0^2 c$. The result reads
\begin{equation} \label{eq:Lmass_load3}
 \lambda_{\rm ML,ph} 
   =  1.32 \cdot 10^{16} \frac{V_{\rm wind}}{c} \frac{R}{\langle l \rangle} 
   	\left(\frac{\eta_X}{10^{-4}} \right)^{-1}
	\frac{n_{\rm N,-4}^{-1}}{ \mathcal{F}(\Gamma) \, \gamma} \, \rm cm \,.
 \end{equation}
It is seen that $\lambda_{\rm ML,ph}$ can be much smaller than $d_0$ due to the fact that its value is inversely proportional to the electron Lorentz factor, whose typical value is $\gamma \approx 10^5-10^6$. Even for Lorentz factor as low as $10^3$, a value estimated from the modelling of some pulsars \cite[see, e.g.,][]{Dubus15}, $\lambda_{\rm ML,ph}$ remains smaller than $d_0$.
It will be shown in \S\ref{sec:non-rel-model} that Eq.(\ref{eq:Lmass_load2}) is the correct definition of the length scale that determines whether mass loading will influence the dynamics of a \emph{non-relativistic} wind. On the other hand, for a \emph{relativistic} wind the definition of $\lambda_{\rm ML} $ has to be modified as it becomes necessary to take into account the relativistic inertia of the wind. We will discuss this issue in \S\ref{sec:rel-model} where we will derive the relativistic version of Eq.(\ref{eq:Lmass_load2}) using a relativistic HD model. Moreover, we will show that there is no significant difference between the relativistic and non-relativistic results, apart from the different definition of $\lambda_{\rm ML}$.

\subsection{Collisional ionization inside the pulsar wind} \label{sec:collisions}
The collisional ionization length scale depends on the Bethe cross section \citep{Kim00}
\begin{equation} \label{eq:sigma_Bethe}
 \sigma_{\rm Bethe}(\gamma) = \frac{4 \pi a_0^2 \alpha^2}{\beta^2} \left[ M^2 \left( \ln(\gamma^2 \beta^2) - \beta^2\right) + C_R\right] \,,
\end{equation}
where $\gamma$ is the Lorentz factor of the electrons (and positrons) and $\beta= (1-\gamma^{-2})^{1/2}$.  The two constants $M^2$ and $C_R$ are related to the atomic form factors of the target, and are independent of the incoming particle energy.  For hydrogen atoms $M^2=0.3013$ and $C_R=4.3019$ \citep{Kim00}.  Using the expression above, together with the typical value $\gamma= \gamma_5 10^5$, furnishes the collisional ionization mean free path in the pulsar wind:
\begin{equation} \label{eq:Lion_wind}
 \lambda_{\rm col} = \frac{V_{\rm NS}}{n_e  \sigma_{\rm Bethe}(\gamma)\, c } 
   = 3.86 \times 10^{23} \,V_{300}^3 \, n_{\rm ISM,-1} \, \gamma_5 \, {\rm cm}\,.
 \end{equation}
As discussed in the previous section, the electron density of the wind is estimated by equating the luminosity of the pulsar with the energy flux in the shocked wind, \emph{i.e.}, $\dot{M} \equiv \mathcal{L}_{w}/(\gamma c^2) = n_e m_e \pi d_0^2 c$ (thereby ensuring that the dynamics of the pulsar wind is not dominated by the {\Bf}).  Using the same values for the parameters as those used in Eq.(\ref{eq:d0}) leads to the estimate $n_{e}\approx 8 \cdot10^{-9} \, \gamma_5^{-1} \, n_{\rm ISM,-1}\, \text{cm}^{-3}$. Note that in the relativistic regime ($\gamma \gg 1$) $\sigma_{\rm Bethe}$ increases only logarithmically with energy, and that increasing $\gamma$ from $10^3$ to $10^6$ implies a decrease in $\lambda_{\rm col}$ of only $\sim40\%$.  

Similar to the case of photo-ionization, (see the discussion after Eq.(\ref{eq:L_ph_ion})) we define $\lambda_{\rm ML, col}$ as the typical mass-loading scale where the loaded mass density is equal to the initial mass density  of the wind, \emph{i.e.}, 
\begin{equation} \label{eq:Lmass_load}
 \lambda_{\rm ML, col} 
   =  \frac {\rho_e}{\rho_N} \frac{V_{\rm wind}}{n_e  \sigma_{\rm Bethe}(\gamma)\, c } 
   =  2.7 \times 10^{19} \, \frac{V_{\rm wind}}{c} 
       \frac{1}{n_{\rm N,-4}} {\rm cm}  \,.
 \end{equation}
It is again stressed that the above expression is for a non-relativistic wind and that the correct expression for the enthalpy should be used for a relativistic wind. Comparison of Eq.(\ref{eq:Lmass_load}) and Eq.(\ref{eq:Lmass_load3}) shows that photo-ionization is considerably more effective than collisional ionization, even for very low luminosity nebulae. Collisional ionization can thus be safely neglected in all realistic cases.

\subsection{Ionization inside the free wind region}
\label{sec:rel_wind}
There is a finite probability that neutrals can be ionized inside the region of the unshocked wind (shaded region in Fig.(\ref{fig:sketch1})). These ions could change the wind dynamics but, as we argue below, this process can be neglected. 
In fact, for parameters typical to bow shock nebulae, the Larmor radius of these ions, calculated in the rest frame of the wind, is comparable or larger than the size of the unshocked wind region. 

To show that this is, indeed, the case, an expression relating the termination shock radius and the wind luminosity is required. Assuming that the energy fluxes of the \Bf\ and particles are similar, the wind luminosity can be estimated as $\mathcal{L}_w = B^2 R_t^2 c$, where $R_t$ is the radius of the termination shock and $B$ is the magnetic field strength at the termination shock. In the rest frame of the wind the newly formed ion have a bulk Lorentz factor equal to the wind Lorentz factor, $\gamma_w$, hence the Larmor radius is:
\begin{equation} \label{eq:fw1}
  r_L=  \frac{\gamma_w m_p c^2}{e B} = \frac{\gamma_w m_p c^{5/2} R_t}{\sqrt{\mathcal{L}_w}} \,,
\end{equation}
and the ratio between $r_L$ and the size of the free wind region can be expressed as
\begin{equation} \label{eq:fw2}
  \frac{r_L}{R_t} = \frac{m_p c^{5/2} \gamma_w}{\sqrt{\mathcal{L}_w}} \approx 5 \, \gamma_6 \, \mathcal{L}_{w,34}^{-1/2} \,.
 \end{equation}
The ionization of neutrals in the unshocked wind can produce ions with $r_L > R_t$ that will escape the free wind region, suffering only a small deflection. The requirement for the production of ions in the free wind regions is a high bulk Lorentz factor and a low spin-down power $\mathcal{L}_w$.  Consequently we neglect the role of these ions and we will only account for neutrals ionized in the shocked wind.

\section{Hydrodynamic model for non-relativistic wind}
 \label{sec:non-rel-model}

\begin{figure}
\begin{center}
\includegraphics[width=0.4\textwidth]{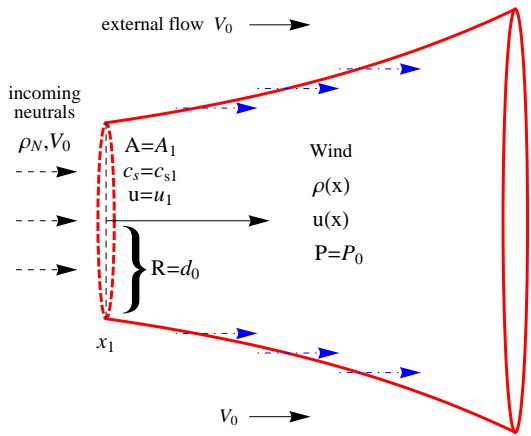}
\end{center}
\caption{Sketch of the simplified model used to study the effect of mass loading in the tail of the bow shock nebula. This Figure represents a zoom in of the dot-dashed rectangle of Fig.~(\ref{fig:sketch1}).  Note that the presence of the bow shock is neglected. The dash-dotted arrows show the regions where the neutrals in scenario B penetrate into the wind from the side of the contact discontinuity.  This is in contrast to scenario A where neutrals only penetrate into the wind in a cylindrical region with cross section $A_1$.}
\label{fig:sketch2}
\end{figure}

To study the effect of mass loading  in the tails of bow shock nebulae, the steady-state conservation equations for mass, momentum and energy are solved in a quasi 1-D approximation. Fig.~(\ref{fig:sketch2}) shows a sketch of the model and represents an idealization of the region enclose in the dot-dashed rectangle of Fig.~(\ref{fig:sketch1}).  The quasi 1-D approximation implies that the transverse cross section, $A$, of the flow can change, but that all the characteristic quantities of the wind, \emph{i.e.}, the velocity, $u$, density, $\rho$ and pressure, $P$, are assumed to be a function of the position $x$ only.  This approach further implies that any internal structures are neglected, in particular the free wind region and the termination shock shown in Fig.(\ref{fig:sketch1}).  For the sake of simplicity the presence of the bow shock is also neglected, but note that possible effects of mass loading on the shape of the bow shock will be discussed in \S\ref{sec:limits-NR}.  It is further assumed that the external medium has a spatially independent velocity, $V_0$, pressure, $P_0$, and density, $\rho_0$.  Lastly, it is assumed that both the ISM and the wind are non-relativistic with an adiabatic index $\gamma_g=5/3$. In the next section we will develop the model for a wind with a relativistic temperature moving with a non relativistic bulk speed. The latter case is the more realistic scenario for a pulsar wind, whereas the former case applies more to stellar winds\footnote{For stellar winds it is necessary to also take into account the charge-exchange that occurs between neutral hydrogen coming from the ISM and protons inside the wind.}. We anticipate that the main difference between these two cases lies solely in the different values used for the enthalpy.  While this changes the dynamical length-scale significantly, it does not change the qualitative behavior of the solution. Finally, note that the steady-state nature of the system is not guaranteed. In fact, we will show that scenarios exist where the steady-state assumption is most likely violated.  

The conservation equations for mass, momentum and energy for a quasi 1-D system, written in the rest frame of the pulsar, are
\begin{align} 
 \partial_x \left[ \rho u A \right] & = q A' \,,   \label{eq:flux_mass} \\
 \partial_x \left[ \rho u^2 A \right] + A\partial_x P  & = q A' V_0 \,,  \label{eq:flux_mom} \\
 \partial_x \left[ \left( \frac{1}{2} \rho u^2 + \frac{\gamma_g}{\gamma_g-1}P \right) u A \right] & = q A' V_0^2/2\,.  \label{eq:flux_en}
\end{align}
Note that $\rho=\rho_e+\rho_p$ is the total density of the wind and that the quantities on the right-hand side of Eqs.(\ref{eq:flux_mass})-(\ref{eq:flux_en}) represent the mass, momentum and energy flux due to the ionization of neutrals, respectively. The mass loading term is given by
\begin{equation}
 q= \dot{n} (m_e+m_p) \,. 
\label{eq:q}
\end{equation}
As we showed in \S\ref{sec:ph-ionization}, $\dot n$ is determined predominantly by photo-ionization, allowing one to write
\begin{equation} 
  \dot n= n_{N} n_{ph} \bar\sigma_{\rm ph} c  \,,
\label{eq:dot_n}
\end{equation}
where $n_N$ is the local density of neutrals, $n_{ph}$ is the photon density as seen in the rest frame of neutrals, and $\bar\sigma_{\rm ph}$ is the photo-ionization cross section averaged over the photon distribution. For the sake of simplicity we assume that both $n_{N}$ and $n_{ph}$ are constant along $x$. The spatial independence of $n_{N}$ is a good approximation when $x \ll \lambda_{\rm ph}$ (\emph{i.e.}, the photo-ionization length scale given in Eq.(\ref{eq:L_ph_ion})), while it becomes necessary to include the spatial dependence of $n_N$ when $x \gtrsim \lambda_{ph}$. This can be done with a simple change of variables as we show in Appendix \ref{app:A}. Conversely, a correct evaluation of the spatial dependence of the photon density is non-trivial and requires a detailed analysis of the specific object one wants to study. Here we avoid such a complication by assuming a constant value.

It is also assumed that the incoming neutrals have the same temperature as the ISM ($\approx 10^4$ K), and they can thus be considered as being cold with respect to the wind in the nebula. As a result of this assumption, the momentum and energy injection terms on the right-hand side of Eq.(\ref{eq:flux_mom}) and (\ref{eq:flux_en}) only contain the contribution due to the bulk motion.

The variable $A'$ represents the effective area crossed by neutrals. In \S\ref{sec:loading} it was noted that neutrals can penetrate into the wind by crossing the head of the nebula.  It is thus required that the effective area should be $A'\approx A_1=\pi d_0^2$. However, in reality the situation is more complicated. We also noted that the CE process produces a diffusion of neutrals in the shocked ISM and that this can lead to neutrals penetrating into the wind from the side of the contact discontinuity. A rough estimate of this process, considering a constant diffusion coefficient, would result in a mass loading that is proportional to $A(x)^{1/2}$.
Moreover, it will subsequently be shown that the effect of mass loading is to expand the cross section of the wind.  When this happens the distance between the bow shock and the contact discontinuity is reduced, thereby further increasing the probability of neutrals penetrating from the side of the contact discontinuity, specifically in the tail region of the pulsar wind.  This results in a mass loading that is proportional to the total area $A(x)$.  A realistic solution therefore requires a description of the bow shock geometry and how the neutrals interact in the shocked ISM.  Such a complication necessitates the use of a 2-D simulation, and is beyond the scope of the present work. In this study the two opposite situations, $A' = A_1$ (scenario A) and $A' \propto A$ (scenario B), are investigated as we expect a realistic situation to be bracketed between these two scenarios.

A consequence of the steady-state assumption is that the pressure is required to be constant everywhere, \emph{i.e.}, $P=P_0$ and $\partial_x P=0$, and it is thus possible to simplify Eqs.(\ref{eq:flux_mom}) and (\ref{eq:flux_en}).   

As a first step to finding the solution, Eq.~(\ref{eq:flux_mass}) can be substituted into the right-hand side of Eqs.~(\ref{eq:flux_mom}) and (\ref{eq:flux_en}), leading to 
\begin{align} 
 \partial_x \left[ \rho u A(u-V_0) \right] & = 0 \,,   \label{eq:const1} \\
 \partial_x \left[ \frac{1}{2} \rho u A \left( u^2-V_0^2 \right) + \frac{\gamma_g P_0 u A}{\gamma_g-1}  \right] & = 0 \,.  \label{eq:const2}
\end{align}
These equations define two constants of the system which can be used to write down two expressions, one for $\rho$ and the other for $A$, as functions of the velocity $u$ only. Evaluating the constants of the system at the position $x=x_1=0$, where the boundary conditions are defined as $u=u_1$, $A=A_1$ and $\rho=\rho_1$, the following expression for the area is obtained: 
\begin{equation} \label{eq:A_u}
 \frac{A(u)}{A_1} = \frac{u_1}{u} +\frac{(u_1-u)(u_1-V_0)(\gamma_g-1) M_1^2}{2 u u_1} \,,
\end{equation}
and the following expression for the density:
\begin{equation} \label{eq:rho_u}
\frac{\rho(u)}{\rho_1}= \left[  \frac{u - V_0}{u_1-V_0}
				 + \frac{(u_1-u)(u - V_0)(\gamma_g-1) M_1^2}{2 u_1^2}  \right]^{-1} \,,
\end{equation}
where $M_1=u_1/c_{s1}$ is the initial Mach number of the wind and $c_s=\sqrt{\gamma_g P/\rho}$ is the sound speed. Note that all quantities with the subscript $_1$ refer to values at the boundary $x_1$. 
 
The next step requires finding an expression for the velocity. Differentiating Eq.~(\ref{eq:flux_mom}) by parts, and using Eq.~(\ref{eq:flux_mass}) leads to
\begin{equation} \label{eq:du_dx}
\partial_x u = - q \frac{A' (u - V_0)}{\rho u A} \,.
\end{equation}
This differential equation can easily be integrated by parts from $x_1=0$ to $x$, after the quantity $\rho u A$ has been substituted using Eq.(\ref{eq:const1}). The solution for scenario A, where $A'=A_1$, is straightforward and reads:
\begin{equation} \label{eq:u(x)_nrA}
 u(x)  = \frac{\lambda u_1 + x V_0}{\lambda + x}  \,,
\end{equation}
where we have introduced the length scale $\lambda=u_1 \rho_1/q$. If $q$ is calculated using the photo-ionization cross section, $\lambda$ corresponds exactly to the definition provided in Eq.(\ref{eq:Lmass_load2}). An important and noteworthy result is that the solution (\ref{eq:u(x)_nrA}) does not depend on the initial Mach number of the wind. Eq.(\ref{eq:u(x)_nrA}) further shows that $u(x)$ decreases monotonically from $u(0)= u_1$ to $u(\infty) = V_0$, and that the typical length scale for this transition  is $\lambda u_1/V_0$. 
The solution for scenario B (\emph{i.e.}, $A'=A$) is slightly more complicated but can be expressed in an implicit form as follows:
\begin{equation} \label{eq:u(x)_nrB}
 \frac{x}{\lambda} = F_1 \frac{u_1-u}{u-V_0} + F_2 \ln \left[ \frac{u_1 - V_0}{u-V_0} \frac{2 u_1^2 + \alpha (u_1-u)}{2 u_1^2} \right] \,,  
\end{equation}
where 
\begin{align} 
  F_1 =  \frac{2 u_1 V_0}{2 u_1^2 + \alpha (u_1-V_0)}  \,, \\
  F_2 = \frac{2 u_1^2 (u_1- V_0)(2 u_1+\alpha)}{ \left[ 2 u_1^2 + \alpha (u_1-V_0) \right]^2}  \,,
\end{align} 
and
\begin{equation} \label{eq:alpha}
  \alpha= M_1^2 (u_1-V_0) (\gamma_g-1) \,.
\end{equation}

The solution (\ref{eq:u(x)_nrB}) also shows that $u(\infty) = V_0$.  This can be readily understood: when the amount of mass loaded is much larger then the initial mass of the wind, the final state of the wind will be the same as the initial state of the neutrals.  This is confirmed by the behaviour of $c_s$, which is also a monotonic decreasing function of $x$, as well as by the result $u\rightarrow V_0 \Rightarrow c_s(u) \rightarrow  0$, which one expects from the assumption of cold neutrals. The continual process of mass loading therefore decelerates the wind from $u_1$ to $V_0$. The main difference between solutions (\ref{eq:u(x)_nrA}) and (\ref{eq:u(x)_nrB}) is the typical length scale where this transition occurs: in the former case it is $\lambda_A \sim \lambda u_1/V_0$, whereas in the latter case it is $\lambda_B \sim \lambda$. This result once again has a simple explanation: the amount of mass loaded in scenario A is a factor $A_1/A$ smaller compared to the amount of mass loaded in scenario B.  For $x\rightarrow \infty$ the solution (\ref{eq:A_u}) shows that the cross section of the flow increases asymptotically to the value
\begin{equation} \label{eq:A_infty}
 \lim_{x\rightarrow\infty} A(u) = A_1 \, \frac{u_1}{V_0}  \left(1+\frac{\gamma_g-1}{2}\, M_1^2 \right) \,,
\end{equation}
where this asymptotic value is the same for both scenarios A and B. As we expect the transition length scale to be proportional to the loaded mass, we have $\lambda_A \simeq \lambda_B A_{\infty}/A_1 \simeq \lambda_B u_1/V_0$.  

\begin{figure}
\begin{center}
\includegraphics[width=0.47\textwidth]{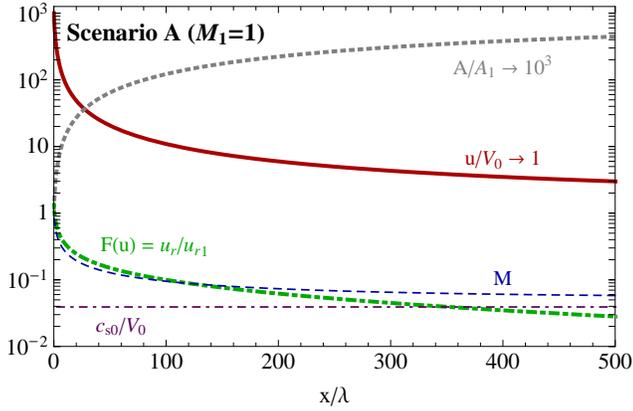}
\end{center}
\caption{Structure of the wind when mass loading occurs in scenario A, as calculated using the non-relativistic model.  The plot shows, as a function of the position, the wind velocity divided by $V_0$ (solid line), the Mach number, $M$ (dashed thin line), and the expansion velocity $u_r$ normalised to $u_{r1}$, as given by Eq.(\ref{eq:u_r1}) (dot-dashed line).  The plot also shows the normalised area, $A/A_1$ (dotted line) and the ISM sound speed normalized to $V_0$ (thin dot-dashed line). The initial wind velocity is $u_1/V_0=1000$ and the initial Mach number of the wind is $M_1=1$.  Note that the spatial coordinate, $x$, is normalised to the mass loading length scale $\lambda$.}
\label{fig:velocity-scenA}
\end{figure}

\begin{figure}
\begin{center}
\includegraphics[width=0.47\textwidth]{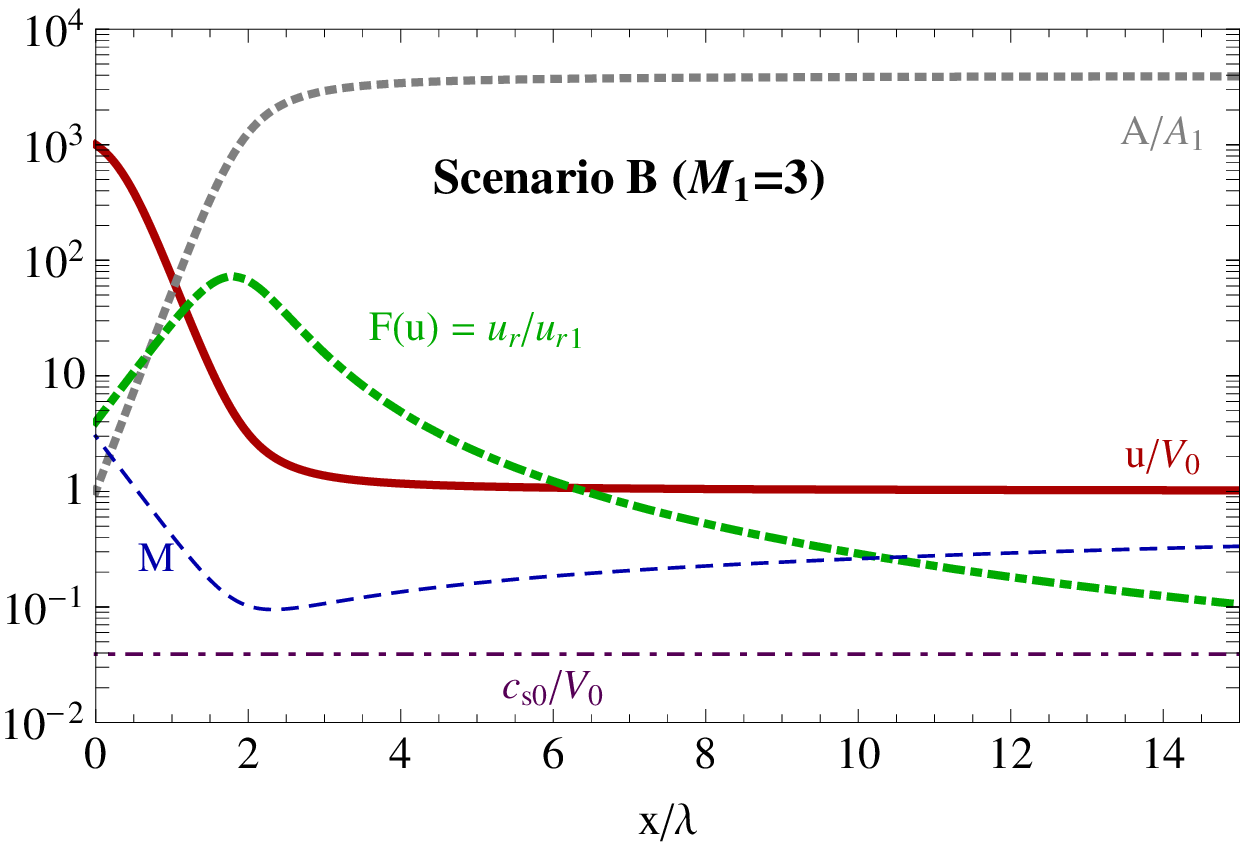}
\includegraphics[width=0.47\textwidth]{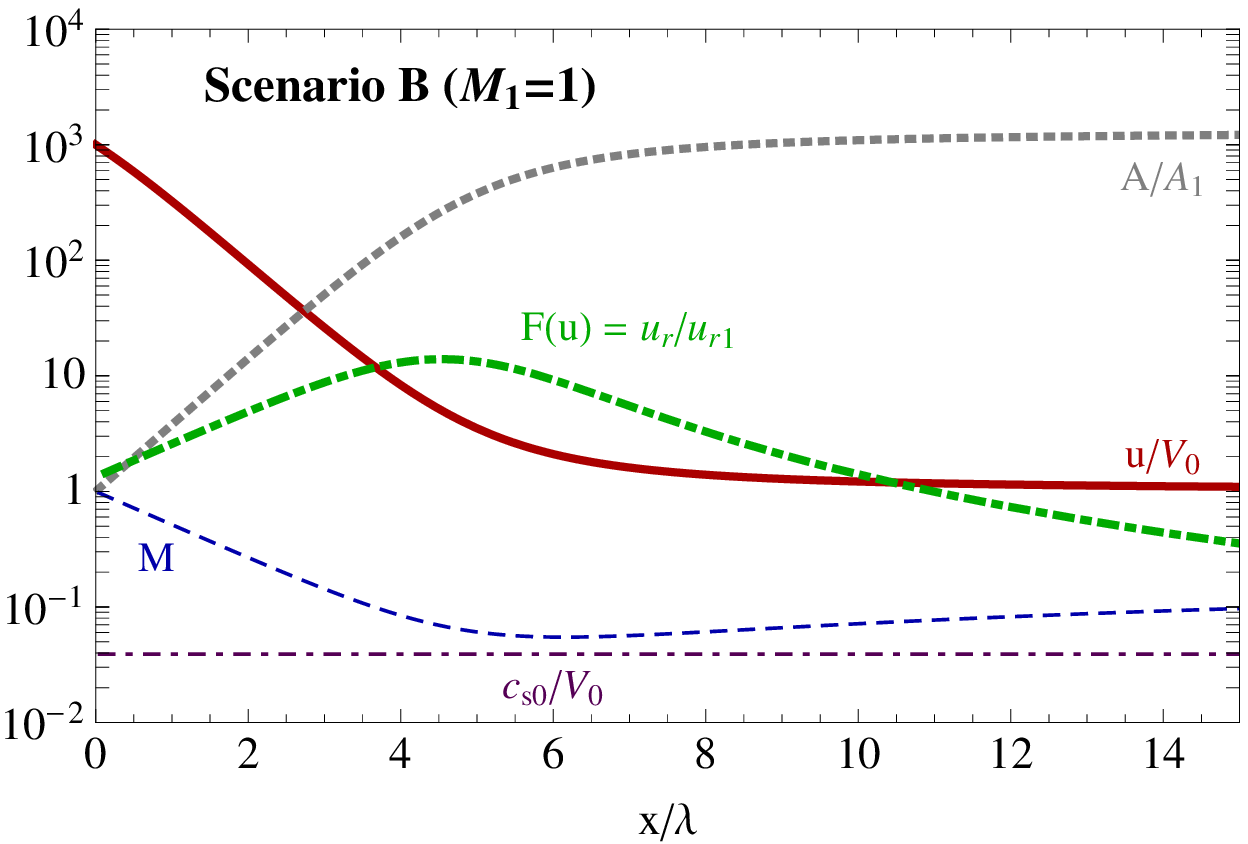}
\includegraphics[width=0.47\textwidth]{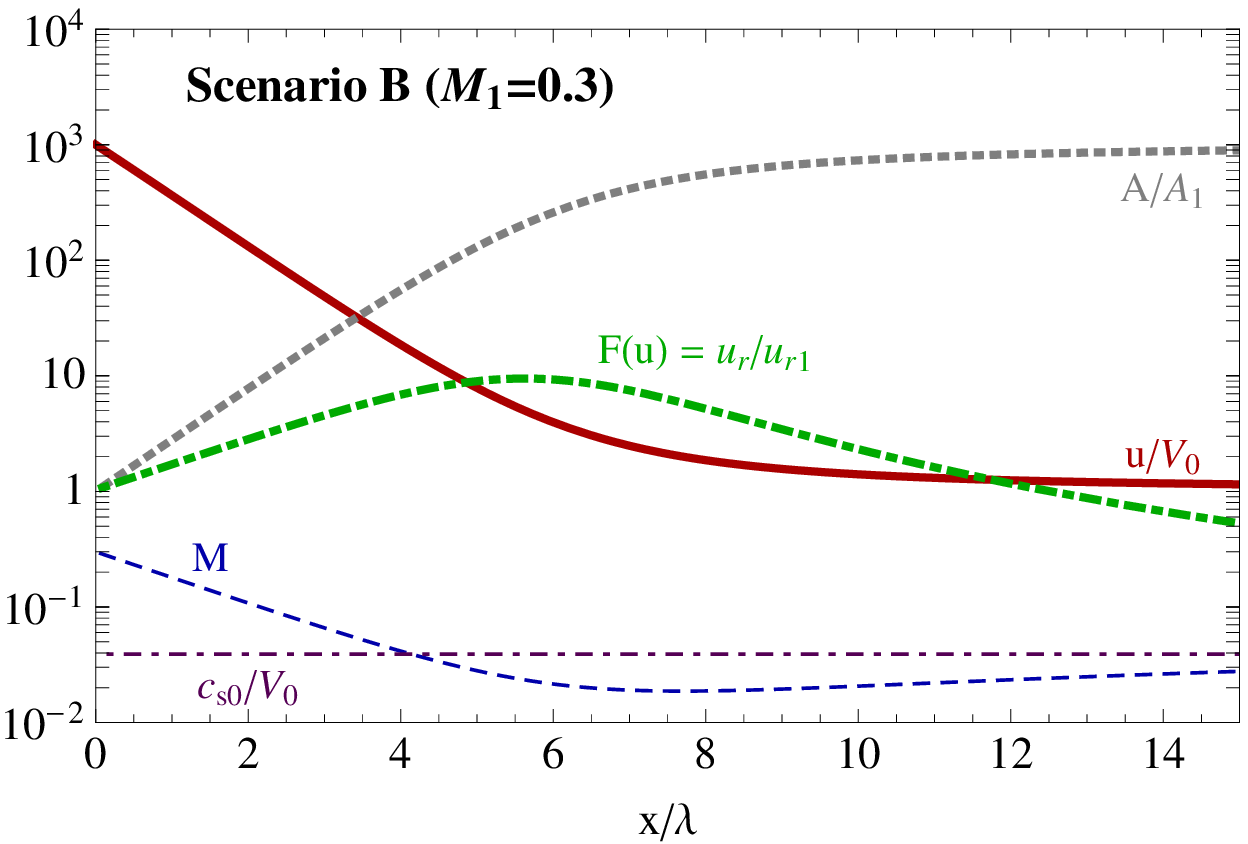}
\end{center}
\caption{The same as Fig.(\ref{fig:velocity-scenA}) but for scenario B. From top to bottom: solutions are calculated for the initial Mach number equal to 3, 1 and 0.3, respectively, while the initial wind velocity is identical for all panels, \emph{i.e.}, $u_1/V_0=1000$.}
\label{fig:velocity-scenB}
\end{figure}

The solutions for $u(x)$, $M(x)$, and $A(x)$ corresponding to scenario A are shown in Fig.~(\ref{fig:velocity-scenA}), and the solutions corresponding to scenario B in Fig.~(\ref{fig:velocity-scenB}).  For the latter case three panels are presented showing the results for three different values of the Mach number, while for scenario A only the case $M_1=1$ is shown as the solution depends only weakly on the value of $M_1$. All these solutions are obtained by using the benchmark values summarised in Table \ref{tab:1}. These values lead to an asymptotic expansion of $A_{\infty}/A_1\approx 1000$,  which, in turn, translates into a radial expansion of a factor $\sim 30$.

Comparison of Figs.~(\ref{fig:velocity-scenA}) and (\ref{fig:velocity-scenB}) confirms that the transition in scenario B occurs much faster than in scenario A. However, apart from the different scale lengths of the transition, the two scenarios are very similar. Fig.~(\ref{fig:velocity-scenB}) further shows that when the initial Mach number increases from 0.3 to 3, the transition occurs faster by a factor $\sim 3.5$. 

A salient feature of Figs.~(\ref{fig:velocity-scenA}) and (\ref{fig:velocity-scenB}) is that there is no qualitative difference between the solutions of a supersonic and a subsonic wind.  This is a remarkable result given that the state of the wind downstream of the termination shock is not well known: the post-shock wind in the head of the nebula, more specifically in the nose region between the termination shock and the contact discontinuity, is re-accelerated to supersonic speeds, while the wind crossing the backward spherical shock is subsonic \cite[see e.g.][]{Bucciantini05}. The average state of the wind is thus not well defined. 
This result seems to be at odds with mass loading models developed in a pure 1-D geometry. \cite[see \eg][\S2, for a review]{Szego00}. More specifically, the solution in the 1-D model generally predicts that mass loading in a supersonic flow leads to the deceleration and heating of the
flow, while mass loading in a subsonic flow causes acceleration and cooling. The reason why this behaviour is not observed in the present model is due to the fact that the flow in a quasi 1-D model can expand in the radial direction while the pressure remains constant, and therefore no acceleration in the flow direction is possible. On the other hand, if the radial expansion were limited for some reason (\eg\  due to hoop stresses caused by to a toroidal magnetic field), then an acceleration of the flow would be possible.  This specific point will be developed in more detail in a subsequent paper.

\begin{table}
\caption{\label{tab:1} Summary of the parameter values used for scenarios A and B. }
\begin{center}
\begin{tabular}{ccccccc}
\hline
 $ n_{\rm ISM}$ & $X_{\rm ion}$ & $T_{\rm ISM}$ &    $V_0$      & $u_1$ & $M_1$ & $\mathcal{L}_{w}$  \\
  $\rm [cm^{-3}]$    &                        &        [ K ]           &    [$\text{km}\,\text{s}^{-1}$]       &            &             & [$\text{erg}\,\text{s}^{-1}$]     \\
\hline
\hline
 0.1  & 0.9 & $10^4$  & 300 & $c$ & 1 & $10^{34}$ \\
\hline
\end{tabular}
\end{center}
\end{table}

\section{Possible implications of mass loading for the bow shock} \label{sec:limits-NR}

\begin{figure}
\begin{center}
\includegraphics[width=0.47\textwidth]{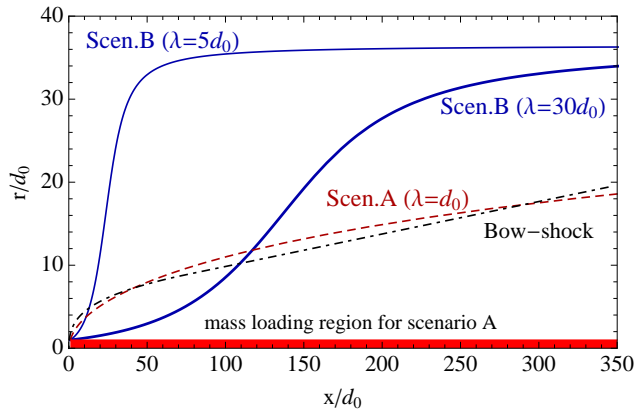}
\end{center}
\caption{Comparison between three mass loaded wind profiles, calculated using the non-relativistic model (solid lines for scenario B and dashed line for scenario A), and the profile of the unperturbed bow shock, calculated using the thin shock approximation (dot-dashed line) \citep{Wilkin96}. The results for scenario B are shown for two different values of the mass loading distance, $\lambda= 5 d_0$ (thin lines) and $\lambda= 30 d_0$ (thick lines), while scenario A is plotted only for $\lambda=d_0$. All cases have an initial Mach number $M_1=1$. The shaded region, with radius $r=d_0$, shows the mass loading region for scenario A.}
\label{fig:bow_shock-NR}
\end{figure}

One of the limitations of the present, quasi 1-D model is that the bow shock which separates the unshocked ISM from the shocked ISM is neglected.  It is therefore useful to compare the results of the present model with the expected structure of the bow shock when mass loading is absent.  In Fig.~(\ref{fig:bow_shock-NR}) the expansion profiles for scenarios A and B presented in the previous section are compared with the profile of an ideal bow shock. The bow shock profile is calculated using the thin-shock approximation \cite[see][Eq.(9)]{Wilkin96} close to the head of the nebula. When the distance from the head becomes large, the thin-shock approximation is no longer valid, and we therefore replace this solution with the Mach cone, where the inclination angle, $\theta$, between the bow shock and the pulsar velocity is such that $\sin \theta = 1/M_{\rm NS}$. For the bow shock profile the benchmark values summarised in Table \ref{tab:1} are again used. 
For the mass loaded wind profiles three different cases are plotted: scenario A with $\lambda/d_0=1$, and scenario B with $\lambda/d_0=5$ and $\lambda/d_0=30$.  Fig.~(\ref{fig:bow_shock-NR}) shows that for all mass loaded cases the pulsar wind in the tail of the nebula expands beyond the position of the unperturbed bow shock, and one would thus expect the geometry of the bow shock to be affected. However, while the expanding wind profile in scenario A closely follows the profile of the bow shock, the expansion of the wind profile in scenario B is much faster, and is capable of producing the head-shoulder shape observed in some H$\alpha$ bow shock nebulae.  It should also be noted that the wind profile in scenario A never crosses the bow shock when $\lambda \gtrsim d_0$, and one would therefore not expect such a configuration to affect the bow shock profile.

\begin{figure}
\begin{center}
\includegraphics[width=0.47\textwidth]{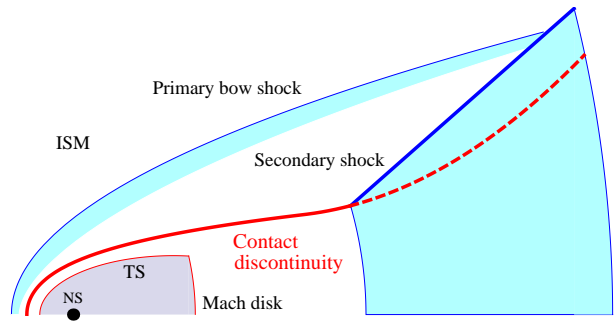}
\end{center}
\caption{Sketch of the generation of a secondary shock inside the first bow shock due to the mass-loading induced expansion of the wind. The shaded region behind the secondary shock shows where one would expect an enhancement of the H$\alpha$ emission.}
\label{fig:secondary_shock}
\end{figure}

We note that neglecting the primary bow shock (\emph{i.e.,} neglecting the ram pressure of the ISM) leads to an incorrect prediction for the wind profile when this profile increases beyond the Mach cone. We show in Appendix \ref{app:B} that when this occurs the formation of a secondary shock in the region between the primary bow shock and the contact discontinuity is expected.  Such a situation is schematically shown in Fig.(\ref{fig:secondary_shock}). Behind the secondary shock the pressure will increase, resulting in a bending of the contact discontinuity towards the axes of the wind. Consequently the rapid expansion predicted in scenario B will probably be attenuated by the presence of the secondary shock.

An interesting consequence related to the presence of the secondary shock is that the increase in temperature will lead to an enhancement of Balmer emission just behind this shock.  However, the enhancement of the emission will strongly depend on the inclination angle of the secondary shock: if the inclination angle is close to the Mach cone, the temperature and Balmer emission will increase only slightly, while for a larger inclination angle the temperature will increase significantly, resulting in a strong enhancement of the Balmer emission.

Although the quasi 1-D model has limitations, it is nevertheless interesting to compare the expansion speed of the pulsar wind with the sound speed of the external ISM.  This expansion speed along the radial coordinate $r$, as seen by an observer co-moving with the external medium, is
\begin{equation} \label{eq:u_r}
 u_r (x) = \frac{dr}{dt} = \frac{V_0}{2 \sqrt{\pi A}} \frac{dA}{dx} =  \frac{V_0}{2 \sqrt{\pi A}} \frac{dA}{du} \frac{du}{dx} \,, 
\end{equation}
where we have used $A=\pi r^2$ and $t=x/V_0$.  The last equality can be used to express $u_r$ in a compact form as a function of $u$ only.  Using $du/dx$ from Eq.(\ref{eq:du_dx}) and calculating $dA/du$ from Eq.(\ref{eq:A_u}), leads to
\begin{equation} \label{eq:u_r_2}
 u_r  =  V_0 \sqrt{\frac{A_1}{4 \pi \lambda^2}} \, F(u) \,,
\end{equation}
where the dimensionless function $F(u)$ reads
\begin{equation} \label{eq:u_rF}
 F(u)  = \frac{A'}{A_1} \, \frac{u_1 (u-V_0)^2}{u^2 (u_1-V_0)} \,
 	    \left( 1 + \frac{\gamma_g-1}{2} M_1^2 \frac{u_1-V_0}{u_1} \right) \,.
\end{equation}
The value of $F$ at $x=0$ is 
\begin{equation} \label{eq:lim_F}
  F(u_1) = \frac{u_1-V_0}{u_1} \,\left( 1 + \frac{\gamma_g-1}{2} M_1^2 \frac{u_1-V_0}{u_1} \right) 
\end{equation}
for both scenarios A and B. In the limit $M_1 \ll1$ and $u_1\gg V_0$ one has $F(u_1) = 1$, and consequently the value of $u_r$ at the origin is 
\begin{equation} \label{eq:u_r1}
 u_{r1}  = V_0 \sqrt{ \frac{A_1}{4 \pi \lambda^2} } = V_0 \frac{d_0}{2 \lambda} \,.
\end{equation}
By contrast, for very large distances one has $\lim_{x \rightarrow \infty} F(u) = 0$. The behavior of the radial expansion speed can be seen in Figs.(\ref{fig:velocity-scenA}) and (\ref{fig:velocity-scenB}) where the normalised speed $u_r(x)/u_{r1}=F(u)$ is shown for all plots. With a simple study of the function $F(u)$ it is easy to demonstrate that $u_r$ is a monotonically decreasing function of $x$ for scenario A, while for scenario B the function $F(u)$ has a peak where $u$ equals
\begin{equation} \label{eq:u_star}
 u_{\rm peak}  = \frac{5 V_0 u_1 (2u_1+\alpha)}{4V_0 \alpha + u_1(2u_1+\alpha)}\,,
\end{equation}
where $\alpha$ is given by Eq.(\ref{eq:alpha}). In the limit $u_1\gg V_0$ one has $u_{\rm peak} \rightarrow 5V_0$, which corresponds to a value of the expansion speed that is equal to 
\begin{equation} \label{eq:u_r_peak}
  u_{r,\rm peak} \equiv u_r(u_{\rm peak}) = u_{r1} \sqrt{\frac{u_1}{V_0}} \frac{16}{25\sqrt{5}} 
  								\left( 1+ \frac{\gamma_g-1}{2}M_1^2 \right)^{3/2} \,.
\end{equation}
In addition, always in the limit $u_1\gg V_0$, the peak of the expansion speed is located at 
\begin{eqnarray} \label{eq:x_peak}
  x_{\rm peak}= \frac{\lambda}{2 + M_1^2(\gamma_g-1)}  \times \hspace{3.5cm}  \nonumber \\
  			\hspace{1.5cm} \times\left\{ \frac{1}{2} + 2 \ln\left[ \frac{u_1 \left( 2+M_1^2 (\gamma_g-1) \right)}{8 V_0} \right] 
			\right\}   \,.
\end{eqnarray}
Using parameter values that are typical for a bow shock PWN leads to $u_{r,\rm peak} \approx (10 - 100) u_{r1}$ and $x_{\rm peak} \approx (2 - 6)\lambda$. Comparison of Fig.~(\ref{fig:velocity-scenB}) and Fig.~(\ref{fig:bow_shock-NR}) shows that for $\lambda\lesssim 100 d_0 $ the expansion speed in scenario B is larger than the sound speed in the ISM when the wind crosses the unperturbed bow shock profile. In other words, when the pulsar wind expands beyond the unperturbed bow shock, the expansion is supersonic and one expects a strong modification of the bow shock. By contrast, for $\lambda\gtrsim 100 d_0 $ the expansion of the wind in the ISM is subsonic, and one therefore expects a less pronounced deformation of the bow shock profile.

From Fig.~(\ref{fig:velocity-scenB}) it should also be noted that for $\lambda\lesssim d_0$, the velocity $u_r$ can be larger than the sound speed in the wind as well as the wind velocity $u$. When the former condition is realised the stationary approach is no longer valid, while the second condition leads to a break down of the quasi 1-D approximation. As a result our model can no longer be used.

Based on the above arguments one may speculate that when $\lambda$ decreases to a value close to $d_0$, the non-stationarity of the problem could give rise to a periodic structure of expanding bubbles, similar to those observed in the Guitar Nebula.  However, addressing this scenario using only analytical models is a complicated matter, requiring the use of full 2-D numerical simulations.

From the present investigation it is difficult to predict which of the scenarios, A or B, is the more realistic one.  Based on the 1-D approach, one can state that when the expansion occurs inside the bow shock, the behaviour is most likely well described by scenario A.  In this scenario neutrals repeatedly undergo charge exchange in the shocked ISM, acquiring the same bulk speed, while also flowing parallel to the contact discontinuity between the shocked ISM and the relativistic wind.
On the other hand, when the wind is close to the position of the unperturbed bow shock (or crosses it) the effective distance between the new bow shock and the contact discontinuity could be small enough to allow a relevant fraction of neutral particles to penetrate into the wind. If this is indeed the case, the expansion speed should increase notably, reaching values close to the value predicted by scenario B.
It should also be noted that far from the head of the nebula the neutral fraction of the ISM should be larger as the ionization due to the UV radiation emitted by the nebula is less effective, and that this should produce a faster transition to scenario B.

\section{Hydrodynamical model for relativistic wind} \label{sec:rel-model}
In this section we present the solution of a mass loaded wind accounting for the fact that the pulsar wind is relativistic, hence we will use a relativistic expression for the enthalpy. This is the only difference with respect to the non-relativistic treatment presented in \S~\ref{sec:non-rel-model}. We will show that the solution is very similar to the non-relativistic one, apart from a different value of the transition length scale $\lambda$.

The relativistic equations for the conservation of particle number, energy and momentum, written in the rest frame of the neutron star and for a 1-D system, are
\begin{align} 
 \partial_x \left[ n_{p,e} u A \right]                            & = \dot n A' \,,  \label{eq:flux0_n} \\
 \partial_x \left[ w \gamma_w u A \right]                  & = q c^2 \gamma_0 A' \,,   \label{eq:flux0_en} \\
 \partial_x \left[ w u^2 A \right] + c^2 A\partial_x P  & = q c^2 \gamma_0 A' V_0 \,.  \label{eq:flux0_mom}
\end{align}
Here $n_{e,p}$ is the numerical density of electrons (protons), $w$ is the total wind enthalpy and $\gamma_w$ and $\gamma_0$ are the Lorentz factors of the wind and the neutrals, respectively. The quantities $q$ and $\dot{n}$ are given by Eqs.(\ref{eq:q}) and (\ref{eq:dot_n}), respectively. We notice that in a relativistic treatment it is more convenient to start with the conservation of the particle number, Eq.(\ref{eq:flux0_n}), rather than with the conservation of mass. 
For the reader's convenience, the derivation of Eqs.(\ref{eq:flux0_n})-(\ref{eq:flux0_mom}) is reported in Appendix \ref{app:C}.

As was the case for the non-relativistic model, we again neglect the free wind region and the termination shock. 
We consider only the pulsar wind after the termination shock where it becomes marginally relativistic with a bulk Lorentz factor $\gamma_w \approx 1$. Furthermore, the neutrals are non-relativistic, hence $\gamma_0 \approx 1$, while the steady-state assumption requires the pressure to be constant everywhere, \emph{i.e.}, $P=P_0$ and $\partial_x P=0$. Using these assumptions, it is thus possible to simplify the system (\ref{eq:flux0_n})--(\ref{eq:flux0_mom}) as follows:
\begin{align} 
 \partial_x \left[ \rho_e u A \right] & = \dot n m_e A' \approx q (m_e/m_p) A'\,,  \label{eq:flux1_n_e} \\
 \partial_x \left[ \rho_p u A \right] & = \dot n m_p A' \approx q A' \,,  \label{eq:flux1_n_p} \\
 \partial_x \left[ w u A \right] & = q c^2 A' \,,   \label{eq:flux1_en} \\
 \partial_x \left[ w u^2 A \right]   & = q c^2 A' V_0 \,.  \label{eq:flux1_mom}
\end{align}
Note that in Eqs.(\ref{eq:flux1_n_e}) and (\ref{eq:flux1_n_p}) we neglect the contribution of the electrons to mass loading, and that the approximation $q= \dot n (m_e+ m_p) \approx \dot n m_p$ is used. In order to close the system (\ref{eq:flux1_n_e})-(\ref{eq:flux1_mom}), an expression for the enthalpy is required.  It is generally believed that the pulsar wind predominantly consists of electron-position pairs with highly relativistic temperatures, and we assume that when mass loading occurs there is no energy transfer between electron and protons. In other words, electrons and protons do not reach a thermal equilibrium, but evolve independently with different temperatures. This implies that the electron gas is always highly relativistic, hence the rest mass contribution to enthalpy can be neglected, \emph{i.e.}, $w_e= \epsilon_e+P_e= 4P_e = 4 P_0$. On the other hand, protons are non-relativistic, hence their thermal energy is always negligible with respect to their rest mass energy, and their enthalpy is $w_p = \rho_p c^2$. The total enthalpy of the wind is thus
\begin{equation} \label{eq:rel-enthalpy}
 w = w_p + w_e = \rho_p c^2 + 4 P_0 \,.
\end{equation}
Using these simplifications allows one to obtain an analytical solution for the wind dynamics by following a procedure similar to the one outlined for the non-relativistic case. Combining Eq.(\ref{eq:flux1_en}) and (\ref{eq:flux1_mom}) furnishes a first constant of motion:
\begin{equation} \label{eq:const1-R}
 u A w (u-V_0) = u A (\rho_p c^2 + 4 P_0) (u-V_0) = {\rm const.}
\end{equation}
Similarly, combining Eq.(\ref{eq:flux1_n_p}) and Eq.(\ref{eq:flux1_en}) furnishes a second constant of motion:
\begin{equation} \label{eq:const2-R}
 u A (w-\rho_p c^2) = 4 u A P_0 = {\rm const.}
\end{equation}
Evaluating the constants at the initial position $x=x_1$, where $u=u_1$ and $A=A_1$, Eq.(\ref{eq:const2-R}) gives the solution for the cross section $A$ as a function of the velocity:
\begin{equation} \label{eq:A(u)_R}
 A(u) = u_1 A_1 / u \,,
\end{equation}
while dividing Eq.(\ref{eq:const1-R}) by Eq.(\ref{eq:const2-R}) gives the solution for the proton density as a function of $u$:
\begin{equation} \label{eq:rho(u)}
 \rho_p(u) = \frac{4 P_0}{c^2} \frac{u_1-u}{u-V_0} \,.
\end{equation}
The electron density can be obtained in a similar way using Eq.(\ref{eq:flux1_n_e}) rather than Eq.(\ref{eq:flux1_n_p}), giving the following result
\begin{equation} \label{eq:rho_e(u)}
 \rho_e(u) =  \rho_{e,0} + \frac{m_e}{m_p} \frac{4 P_0}{c^2} \frac{u_1-u}{u-V_0} \,,
\end{equation}
where $\rho_{e,0}$ is the initial electron density.
One noteworthy point is that the solutions (\ref{eq:A(u)_R}), (\ref{eq:rho(u)}) and (\ref{eq:rho_e(u)}) do not depend on the specific assumption regarding the injection cross section $A'$ (this result is also identical to the non-relativistic case). The solution for the velocity $u(x)$ can be obtained by deriving Eq.(\ref{eq:flux1_mom}) by parts, and using Eq.(\ref{eq:flux1_en}). This leads to the expression
\begin{equation} \label{eq:u(x)}
 \frac{\partial u}{\partial x} = - q c^2 \frac{A'}{A} \frac{u-V_0}{w u} \,.
\end{equation}
As was done for the non-relativistic case, we again distinguish between scenario A ($A'=A_1$) and scenario B ($A'=A$). For scenario B the integration of Eq.(\ref{eq:u(x)}) leads to the following implicit solution:
\begin{equation} \label{eq:u(x)_relB}
 \frac{x}{\lambda_{\rm rel}} = \frac{u}{u-V_0} - \frac{u_1}{u_1-V_0} + \ln \left[ \frac{u_1-V_0}{u-V_0} \right] \,,
\end{equation}
where we have introduced the relativistic length scale 
\begin{equation} \label{eq:lambda_rel}
 \lambda_{\rm rel} = \frac{4 P_0 (u_1-V_0)}{q c^2} \simeq \frac{4 P_0}{\rho_N c^2} \frac{u_1}{n_{\rm ph} \bar \sigma_{\rm ph} c}  \,.
\end{equation}
The second equality results from assuming $V_0\ll u_1\sim c$ and using the photo-ionization rate $n_{\rm ph} \bar \sigma_{\rm ph} c$ in the calculation of the mass loading rate, \emph{i.e.}, $q = m_p n_N n_{\rm ph} \bar \sigma_{\rm ph} c$. Comparison of Eq.(\ref{eq:lambda_rel}) with the definition of the mass loading length scale for the non-relativistic case,  Eq.(\ref{eq:Lmass_load2}), shows that they are identical, with the exception that the quantity $4 P_0/c^2$ replaces the electron mass density. In the steady-state model the internal pressure of the wind is equal to the external pressure, hence $4P_0$ corresponds to the specific enthalpy of the electron-positron wind plasma.
Using Eq.(\ref{eq:L_ph_ion}) from \S\ref{sec:ph-ionization} allows one to find a numerical estimate for $\lambda_{\rm rel}$:
\begin{equation} \label{eq:lambda_rel2}
 \lambda_{\rm rel} \approx 1.2 \cdot10^{13} \,T_4 \frac{R}{\langle l \rangle}\, \frac{u_1}{c}  \, 
 	\eta_{X,-4}^{-1} \, V_{300}^{-2} \, n_{N,-4}^{-1} \mathcal{F}(\Gamma)^{-1} \rm cm \,,
\end{equation}
where $P_0= n_{\rm ISM} K_B T$ and $T= 10^4 T_4$ K have been used. Choosing realistic values for the parameters associated with bow shock nebulae emitting H$\alpha$ shows that $\lambda_{\rm rel}$ can be as large as $10^{15}$ cm, but we stress that this value can vary by orders of magnitude, essentially due to the fact that the values of the neutral density inside the wind and the luminosity of the PWN tail are difficult to estimate. Nevertheless, for any scenario $\lambda_{\rm rel}$ represents the lower limit for the typical expansion length scale. For example, in scenario A a larger expansion length scale is predicted. 

Substituting $A'=A_1$ in Eq.(\ref{eq:u(x)}) leads to the following equation
\begin{equation} \label{eq:u_relA}
 \frac{\partial u}{\partial x} = - \frac{q c^2}{4 P_0} \frac{(u-V_0)^2}{(u_1-V_0) u_1} \,,
\end{equation}
which, once integrated by parts, gives the solution
\begin{equation} \label{eq:u_relA(x)}
 \frac{x}{\lambda_{\rm rel}} = \frac{u_1}{u-V_0} - \frac{u_1}{u_1-V_0} \,.
\end{equation}
In this case the typical expansion length scale is  
\begin{equation} \label{eq:lambda_A}
 \lambda_{\rm rel,A} = \frac{u_1}{V_0} \lambda_{\rm rel} = \frac{4 P_0 (u_1-V_0) u_1}{q c^2 V_0}  \,.
\end{equation}
As $u_1/V_0$ can be as large as $10^3$, the mass loading length scale  in scenario A can reach $10^{18}$ cm. In a realistic case one expects the transition to occur between $\lambda_{\rm rel}$ and $\lambda_{\rm rel,A}$. 

Eq.(\ref{eq:lambda_rel2}) predicts that a density of neutrals as small as $10^{-4}\,\text{cm}^{-3}$ is sufficient to have $\lambda_{\rm rel}$ comparable to the size of the nebula. Thus, for a relativistic pair-plasma wind, a relatively small amount of neutrals can strongly affect the tail flow.

\begin{figure}
\begin{center}
\includegraphics[width=0.47\textwidth]{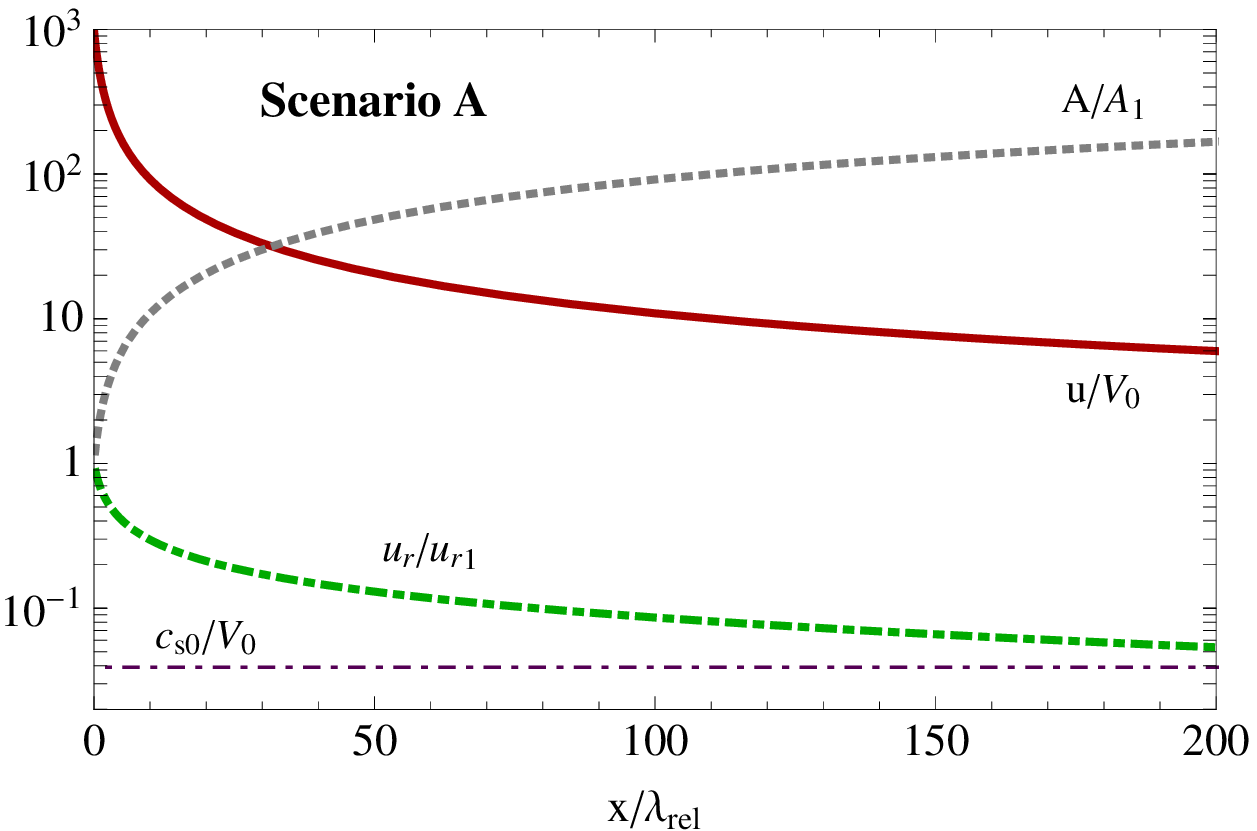}
\includegraphics[width=0.47\textwidth]{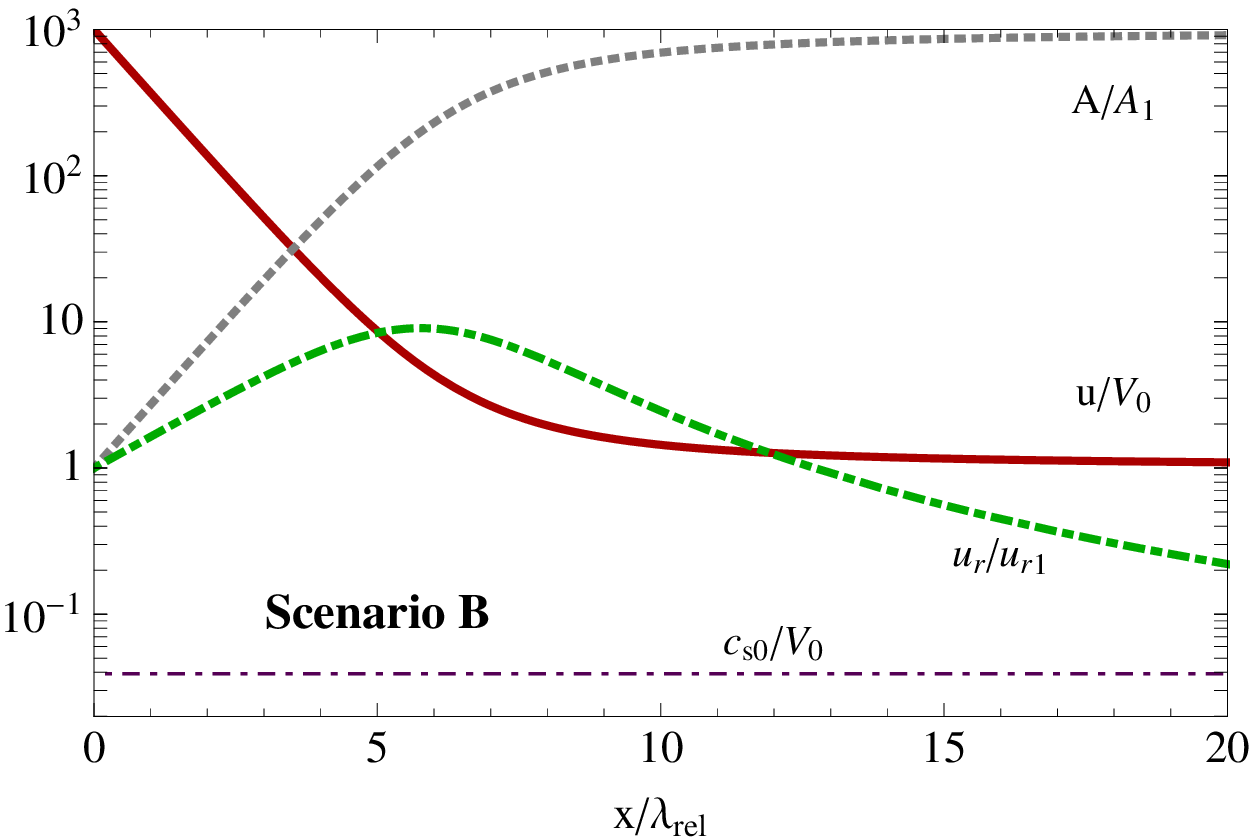}
\end{center}
\caption{Structure of a pulsar wind when mass loading occurs, assuming that the wind consists of a hot relativistic electron-positron pair plasma. Top and bottom panels report results for scenario A and B, respectively, using the same values summarised in Table \ref{tab:1}.  The various lines represent the wind speed divided by $V_0$ (solid line), wind cross section (dot-dashed line) and expansion speed (dotted line) both normalised to their initial values. The sound speed of the ISM, $c_{s0}$, also normalised to $V_0$, is shown for comparison (thin dot-dashed line).}
\label{fig:velocity_rel}
\end{figure}

Fig.(\ref{fig:velocity_rel}) shows the solutions for $u$ and $A$ corresponding to scenarios A and B, where the parameter values given in Table \ref{tab:1} have been used. Comparison of Fig.(\ref{fig:velocity_rel}) with Figs.(\ref{fig:velocity-scenA}) and (\ref{fig:velocity-scenB}) shows that the relativistic and non-relativistic solutions are very similar.

\begin{figure}
\begin{center}
\includegraphics[width=0.47\textwidth]{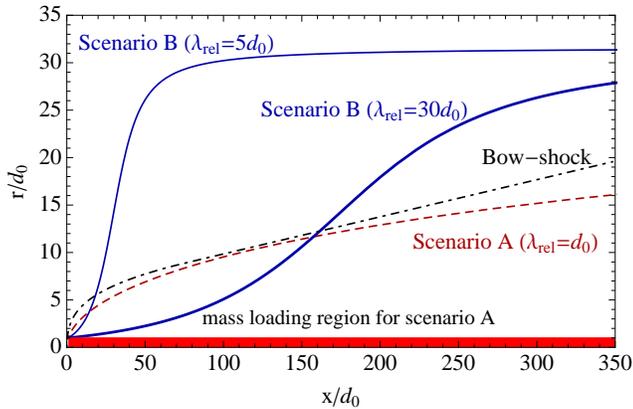}
\end{center}
\caption{Same as Fig.(\ref{fig:bow_shock-NR}), but for a relativistic model. Notice that the dot-dotted line is the same as in Fig.(\ref{fig:bow_shock-NR}) and represents the bow shock profile in the thin shock approximation for non-relativistic wind.}
\label{fig:bow_shock-R}
\end{figure}

As was done for the non-relativistic case, the profile of a relativistic wind undergoing mass loading is compared to the typical profile of the unperturbed bow shock. Fig.~(\ref{fig:bow_shock-R}) is analogous to Fig.~(\ref{fig:bow_shock-NR}), but the wind profile is now calculated using the relativistic expression for scenarios A and B obtained in Eq.(\ref{eq:A(u)_R}). One can see that the relativistic and non-relativistic model give very similar profiles, with the important difference that $\lambda$ is replaced by $\lambda_{\rm rel}$.
Additionally, a similar radial expansion is predicted for the corresponding scenarios of the non-relativistic and relativistic cases. 
Starting from Eq.(\ref{eq:u_r}), using $du/dx$ from Eq.(\ref{eq:u_relA}), and calculating $dA/dx$ from Eq.(\ref{eq:A(u)_R}), leads to the following expression for the expansion speed:
\begin{equation} \label{eq:u_r2}
 u_r  =  V_0 \sqrt{\frac{A_1}{4 \pi \lambda_{\rm rel}^2}} \, \frac{u_1^{1/2} \left(u-V_0 \right)^2 }{u^{5/2}} 
 	    \left( \frac{u}{u_1}\right)^{\frac{1}{2} \pm \frac{1}{2}} \,,
\end{equation}
where the upper and lower signs refer to scenarios A and B, respectively. Fig.~(\ref{fig:velocity_rel}) also shows the normalised velocity $u_r/u_{r1}$, where
\begin{equation} \label{eq:u_r3}
 u_{r1}  = V_0 \left(\frac{A_1}{4 \pi \lambda_{\rm rel}^2} \right)^{1/2} = \frac{d_0}{2 \lambda_{\rm rel}} V_0 \,,
\end{equation}
is the expansion speed at $x=0$, with this speed being the same for both scenarios A and B.  Similar to the non-relativistic case, $u_r(x)$ is also a monotonically decreasing function of $x$ for scenario A, while for scenario B it has a peak at $u=u_{\rm peak}= 5V_0$, corresponding to
\begin{equation} \label{eq:u_r_peak_R}
  u_{r,\rm peak} \equiv u_r(u_{\rm peak}) = u_{r1} \sqrt{\frac{u_1}{V_0}} \frac{16}{25\sqrt{5}} \,.
\end{equation}
The peak is located at the position $x_{\rm peak}$, which, in the limit $u_1 \gg V_0$, is equal to 
\begin{equation} \label{eq:x_peak_R}
  x_{\rm peak}= \lambda_{\rm rel}
  			\left( \frac{1}{4} + \ln\left[ \frac{u_1}{4 V_0} \right] 
			\right)   \,.
\end{equation}
It is interesting to note that this result is identical to the non-relativistic case for the limit $M_1\rightarrow 0$. In other words, all discussions presented for the non-relativistic solutions at the end of \S\ref{sec:limits-NR} also apply to the relativistic solutions. In particular, when $\lambda_{\rm rel} \lesssim d_0$ the expansion velocity can become larger than the wind velocity, thereby causing the quasi 1-D approximation to break down.

\section{A visual fit of bow shock nebula PSR J0742-2822}
 \label{sec:fit} 
In this section we provide an example of a comparison between the predicted profile of the wind due to mass loading and the observed shape of an H$\alpha$ bow shock. Among the pulsars showing an anomalous H$\alpha$  bow shock, we chose PSR J0742-2822 for two reasons:  not only has several quantities been measured with sufficient accuracy, but the pulsar is believed to move almost perpendicular to the line of sight, a fact which, to some extent, simplifies the comparison between the model prediction and observation. 

We use the parameter values reported by \cite{Brownsberger14}. The pulsar luminosity is $\mathcal{L}_w=1.9 \cdot 10^{35}\,\text{erg}\,\text{s}^{-1}$, while the measured proper motion is $29.0$ mas yr$^{-1}$. The most recent distance estimate gives $2.1\pm0.5$ kpc \citep{Janssen06}. Using a value of $d=2$ kpc, \cite{Brownsberger14} estimated $n_{\rm ISM} =0.28\,\text{cm}^{-3}$ for the total ISM density and $V_\perp= 275\,\text{km}\,\text{s}^{-1}$ for the projected component of the pulsar velocity. We assume that the velocity of the pulsar is almost perpendicular to the line of sight, hence $V_0 \approx V_\perp$. 

\begin{figure}
\begin{center}
\includegraphics[width=0.47\textwidth]{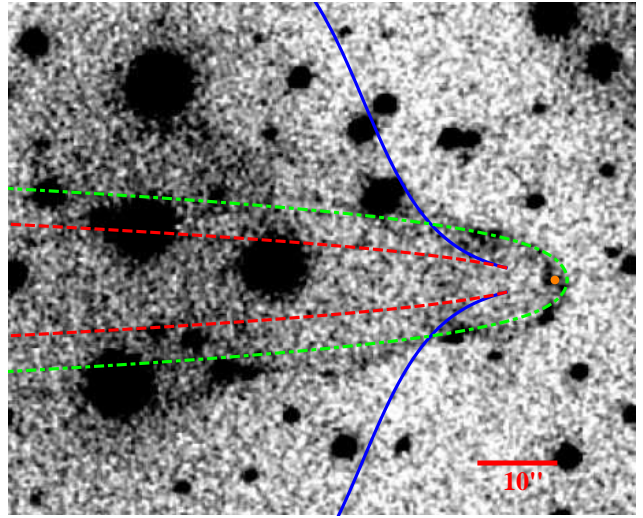}
\end{center}
\caption{Comparison of the observed H$\alpha$ bow shock produced by the PSR J0742-2822 \citep{Brownsberger14} with the wind profile obtained using the relativistic calculations for scenario A (dotted line) and scenario B (solid line). The dot-dashed line shows the predicted position of the unperturbed bow shock. The small dot in the head of the nebula indicates the position of the pulsar. The profile of the wind starts at the location of the backward termination shock, which is $4 d_0$ behind the pulsar.}
\label{fig:J0743Comp}
\end{figure}

Using the above values, the stand-off distance is estimated to be $d_0=3.8 \cdot 10^{16}$ cm. Note that this value is compatible with the measured distance between the pulsar and the apex, measured as $1.4''$, corresponding to a physical distance of $4.2\cdot 10^{16} d/(2 \rm kpc)$ cm. Using this value of $d_0$, the profile of an ideal bow shock (\emph{i.e.,} no mass loading) is calculated according to the procedure outlined at the beginning of \S\ref{sec:limits-NR}.  The resulting solution is shown in Fig.~\ref{fig:J0743Comp} (dot-dashed line), while the small dot indicates the position of the pulsar. The unperturbed bow shock profile fits the head of the nebula reasonably well, but fails to reproduce the ``fan'' structure emerging behind the bow shock head.

For the mass loaded wind, profiles are calculated using the relativistic model developed in \S\ref{sec:rel-model}, with Fig.~\ref{fig:J0743Comp} showing the results for both scenarios A (dashed line) and B (solid line).  In order to ensure that the wind profile in scenario B crosses the bow shock at the precise location where the ``fan'' structure emerges, the profiles are calculated using the value $\lambda_{\rm rel} = 2 d_0$. Note that the wind profiles start at the position $d_{\rm back}=4 d_0$ behind the pulsar, which corresponds to the location of the backward termination shock according to numerical simulations \citep{Bucciantini05}. As previously discussed in \S\ref{sec:limits-NR} and \S\ref{sec:rel-model}, scenario A cannot account for a rapid expanding structure emerging from the unperturbed bow shock as the resulting profile is very smooth, while scenario B does predict such a structure. On the other hand, from Fig.~\ref{fig:J0743Comp} it can be seen that the expansion in scenario B is larger than the one detected in H$\alpha$.

Our aim is not to exactly reproduce the shape of the bow shock as our model only predicts the shape of the relativistic tail wind.  The more important point that we want to emphasise is that the location where the ``fan'' structure emerges is compatible with our predictions.
In particular, one can ask whether the ratio $\lambda_{\rm rel}/d_0 = 2$ is compatible with observations. This can be checked by estimating the neutral density inside the wind. 

Analysing the H$\alpha$ flux, \cite{Brownsberger14} found that the neutral fraction of the ISM is comparable to $\sim 1$. However, as the uncertainty is quite large, we use a fiducial value of 0.5. It was shown in \S\ref{sec:shocked_ISM} that the collisional ionization in the shocked ISM is not effective when $V_0 \lesssim 300\,\text{km}\,\text{s}^{-1}$, therefore one need only take into account the effect of photo-ionization. As was shown in \S\ref{sec:ph-ionization}, the UV flux from the nebula can be estimated from the X-ray luminosity:  the measured upper limit for the X-ray flux is $F_X < 2 \cdot 10^{-14}$ erg cm$^{-2}$ s$^{-1}$ \citep{Abdo2013}, which translates into an upper limit for the X-ray luminosity of $\mathcal{L}_X < 9.6 \cdot 10^{30} (d/2 \rm kpc)^2\,\text{erg}\,\text{s}^{-1}$. For the non-thermal spectrum we assume a power law with an index $\Gamma=2$.  Furthermore, using Eq.(\ref{eq:ion_frac_sol}) allows one estimate the neutral density at the location of the backward termination shock, $d_{\rm back}$. The value of this density averaged over the cross section of the wind is found to be $\sim 0.13 \,n_{\rm ISM}=0.065\,\text{cm}^{-3}$. Finally, using this value in Eq.(\ref{eq:lambda_rel2}) and dividing the result by Eq.(\ref{eq:d0}), one finds that $\lambda_{\rm rel}/d_0 > 0.03 R/\langle l \rangle$.  A value of $\lambda_{\rm rel}/d_0 = 2$ thus implies $R/\langle l \rangle \lesssim 60$, which is fully compatible with the wind geometry.

\section{Mass loading in a magnetised wind}
 \label{sec:magnetic}

In this paper we have neglected the role of the magnetic field inside the wind of the nebula. Nevertheless, using the conservation of magnetic flux, we now show that general conclusions can be drawn from the knowledge of the wind dynamics only.  We consider two different configurations: a purely poloidal and a purely toroidal magnetic field, with both cases assuming that the initial magnetic field is dynamically unimportant. In the poloidal case the flux conservation is applied through the cross section $A$, which leads to $\bar B_x A = const$, where $\bar B_x$ is the average poloidal magnetic field strength. Using the solution for the cross section from Eq.(\ref{eq:A(u)_R}), one finds that $\bar B_x \propto u$. As the fluid speed decreases, the poloidal magnetic field strength will also decreases, and one would not expect the wind dynamics to be affected. 

For the toroidal configuration the opposite situation occurs. The conservation of magnetic flux is applied through a poloidal surface (\emph{i.e.}, a surface parallel to the $x$ direction) which implies $\bar B_\phi u \sqrt{A} = const$, where $\bar B_\phi$ is the average poloidal magnetic field inside the wind. Again using Eq.(\ref{eq:A(u)_R}), one finds $\bar B_\phi \propto u^{-1/2}$. Consequently the strength of the toroidal component increases along $x$. As the toroidal magnetic field exerts a hoop stress on the wind $\propto \bar B_{\phi}^2$, the magnetic pressure can easily become comparable to the internal pressure, thereby reducing the expansion of the wind.

These conclusions have implications for the non-thermal emission. The synchrotron emissivity is proportional to $j_{\rm syn} \propto n_e {\bar B}^2$, where $n_e$ is the density of relativistic electrons and $\bar B$ is the average field. The density of relativistic electrons is constant along $x$, and as a result the expansion induced by mass loading will enhance $j_{\rm syn}$ if the magnetic field is mainly toroidal. Conversely, for a poloidal configuration at the location of expansion the synchrotron emission should decrease. 

Remarkably, this prediction is compatible with the observations of at least one object. The bow shock nebula powered by the pulsar J1509-5850 has been detected both in X-rays and radio, and shows an anti-correlation in these two bands \citep{Hui2007, Kargaltsev08b}: while the head of the nebula is  mainly bright in X-rays, the bulk of the radio emission is observed from the tail of the wind at a distance roughly 4' far away from the pulsar, at a location where the X-ray emission becomes negligible. Moreover, radio polarimetry measurements show that the magnetic field in the tail is mainly toroidal \citep{Ng2010}. 

In our model these observations can be explained by assuming that the enhancement of the radio emission occurs where the tail expands due to mass loading, while the decrease in X-ray emission should be the consequence of radiative cooling, which is efficient for the X-ray emitting electrons but is negligible for the radio emitting electrons. This picture is confirmed by the fact that J1509-5850 has also been observed in H$\alpha$, showing an anomalous bow shock expansion right at the location where the enhancement of the radio emission begins. A further intriguing detail is that the comparison between radio and X-ray data suggests a significant deceleration of the flow when moving downstream, with the speed decreasing by 1-2 orders of magnitude \citep{Ng2010}.  From our model it follows that this decreases can be the result of mass loading.  However, a more quantitative model is needed to better compare this picture with observations.

Noticeably, an opposite situation has been observed for the {\it Mouse} nebula (J1747-2958), where radio polarimetry shows a poloidal magnetic field structure along the tail \citep{Gaensler04}. In this case X-ray and radio emission are both peaked in the head of the nebula and decrease smoothly along the tail. 
The effect of mass loading for the Mouse nebula is probably negligible, in fact the H$\alpha$ emission is not detected, and 
this is probably due to its exceptional luminosity able to photo-ionise the ISM around it. The X-ray luminosity measured by {\it Chandra} is $\mathcal{L}_X=5\cdot10^{34}$ erg s$^{-1}$ for a distance of 5 kpc \citep{Gaensler04},  while the pulsar spin down luminosity is $\mathcal{L}_{w}=2.5\times10^{36}$ erg s$^{-1}$, giving an X-ray efficiency $\eta_x=0.02$. \cite{Gaensler04} also measured the photon index as $1.8<\Gamma_X <2.5$, and provide an estimate of the ISM density of $\approx 0.3$ cm$^{-3}$. Using these values we can estimate the upper limit for the X-ray efficiency from Eq.~(\ref{eq:L_max}) to be $1.2 \cdot 10^{-4}$. Because $\eta_X$ is much larger than this upper limit, we infer that the vast majority of H atoms have been ionized before to reach the bow-shock.

\section{Discussion and conclusions}
 \label{sec:conc}
In this work we have shown that the structure of a bow shock nebula produced by a neutron star propagating through the ISM can be significantly affected by the presence of neutral hydrogen in the ISM.  In many cases a non-negligible fraction of neutral atoms can penetrate into the relativistic wind where they will be ionized by UV photons emitted by the nebula. Once ionized, the new protons and electrons interact with the wind, leading to a net mass loading of the wind in the tail region of the nebula. More specifically, this mass loading is important in the shocked part of the tail wind, while it is most likely negligible in the region of the free, unshocked wind.

To investigate the effect of mass loading, we have developed a steady-state hydrodynamic model using a quasi 1-D approximation, with the study focusing on two specific scenarios. In the first a wind consisting of a non-relativistic plasma with an adiabatic index $\gamma_g=5/3$ (see \S\ref{sec:non-rel-model}) was investigated, while the second situation investigated a wind that consists of a hot relativistic electron-positron plasma, as is the case for a pulsar wind (see \S\ref{sec:rel-model}). Remarkably, both situations show the same qualitative behavior: the loaded mass decelerates the wind and produces a transverse expansion of the tail. The typical expansion factor of the tail cross section is $A_{\infty}/A_1 = u_1/V_0$, where $u_1 \sim c$ is the initial velocity of the pulsar wind after the termination shock, and $V_0$ is the speed of the neutron star measured in the rest frame of the ISM, which is of the order of few hundreds $\text{km}\,\text{s}^{-1}$. The model therefore predicts $A_{\infty}/A_1 \approx 1000$ (for a non-relativistic wind there is a correction due to the initial Mach number of the wind - compare Eq.(\ref{eq:A_infty}) with Eq.(\ref{eq:A(u)_R})). The main difference between the two situations is the distance where this expansion occurs, being determined by the mass-loading length scale, $\lambda_{\rm ML}$.  We further showed that this latter length scale is determined by the distance where the enthalpy of the loaded mass is comparable to the enthalpy of the wind (compare Eq.~(\ref{eq:Lmass_load2}) vs. Eq.~(\ref{eq:lambda_rel})).

Considering a relativistic wind consisting of an electron-positron plasma, and using parameters values observed for pulsar bow shock nebulae, we showed that the mass loading effect could be responsible for the {\it head-shoulder} shape observed in many bow shock nebulae. Unfortunately it is not easy to make a quantitative prediction for specific objects as the amount of mass loading depends on two important parameters, specifically the density of neutrals inside the wind and the density of UV photons, both of which are difficult to estimate. 

We showed that a relatively small density of neutrals inside the wind (as small as $10^{-4}\,\text{cm}^{-3}$) is sufficient to affect the wind, decelerating the tail flow and producing a fast expansion in the transverse direction. For comparison we remember that the typical number density of neutral hydrogen in the warm interstellar medium (where H$\alpha$ bow shock nebulae are though to propagate) is $\sim 0.05-0.5$ cm$^{-3}$  \citep{Jean09}. In order for the mass loading effect to be negligible, either the ISM should be completely ionized, or the photo-ionization due to the nebula should be so effective that all neutrals are ionized before they reach the termination shock in the tail. Neither of these two possibilities can be applied to H$\alpha$ bow shock nebulae as the presence of H$\alpha$ lines is an indication that the photo-ionization is incapable of fully ionizing the ISM.  If this were the case, it would not be possible to observe the H$\alpha$ emission in the first place.

We also made a visual comparison of the wind profile predicted from our model with the H$\alpha$ bow shock observed around PSR J0742-2822. We showed that a density of neutrals in the wind comparable to $\sim 0.06\,\text{cm}^{-3}$ is required to explain the presence and the location of the ``fan'' structure observed behind the head of the nebula. Such a density is compatible with the estimated neutral density of the local ISM, taking into account the photo-ionization resulting from the UV radiation emitted by the nebula, which, in turn, was estimated from the upper limit of the X-ray flux.

Finally, even though the magnetic field is neglected in the present model, we discussed the qualitative behaviour of a magnetised wind flow that is being loaded with mass. The poloidal component of the field decreases with distances, thereby becoming dynamically unimportant. On the other hand, the toroidal component, if present, will be amplified due to the compression of the wind flow. Such an amplification will have two important consequences: the first is to limit the transverse expansion of the wind due to the increase of the magnetic hoop stresses; the second is to produce an enhancement of the synchrotron emission in the location where the expansion occurs. Remarkably, this prediction is confirmed by the observation of the bow shock nebula associated with the pulsar J1509-5850, where an enhancement of the radio emission is observed far from the head of the nebula, and at the same location where the H$\alpha$ emission shows an anomalous expansion of the bow-shock. 

The quasi 1-D approximation presented in this work has the advantage of having a clear and simple analytical solution, thereby making it very usable. However, it also has several limitations. Firstly, we neglected the presence of any internal structure, as well as the bow shock. Moreover, the quasi 1-D approximation is based on the assumption that the transverse expansion velocity is much smaller than the wind velocity, a condition that can be violated when $\lambda_{\rm ML}$ is of the order of, or smaller than the typical stand-off distance of the nebula. In this case the quasi 1-D approximation breaks down. 
Additionally, the transverse expansion speed can be faster than the sound speed of the wind. When this occurs the steady-state condition is violated and the solution presented is no longer applicable.  It is therefore necessary to implement a time-dependent solution. This last situation is especially interesting as it points toward to possibility of having a tail wind with periodic expanding bubbles similar to what is observed in the Guitar nebula. 
Hopefully all this limitations can be overcome using time-dependent, 2-D simulations, which will form the topic of research of a forthcoming paper.

\section*{Acknowledgments}
GM is grateful to Niccol\'o Bucciantini for many fruitful discussions on PWNe and relativistic winds. This work was funded through the NSF grant num. 1306672.

\appendix

\section{Depletion of neutrals inside the wind}
\label{app:A}
For the solutions presented in the paper we have assumed that the density of neutrals inside the wind remains constant, \emph{i.e.}, that the number of ionized atoms is negligible with respect to the total number of neutrals. This is true only on a length scale much smaller than the ionization length scale, $\lambda_{\rm ph}$.  Here we show how to modify the model, should this assumption no longer be valid, by taking into account the fact that the density of neutrals is no longer constant, but decreases as a function of distance.  

When the depletion of neutrals becomes important, scenario B (where neutrals can also penetrate into the wind region through the side of the contact discontinuity) can no longer be described using a 1-D model as the neutral density inside the wind will depend on both $x$ and $r$. The discussion will therefore be limited to scenario A, where the neutrals penetrate only through the head of the nebula.  

It is useful to introduce a new spatial variable, $\xi$, defined as
\begin{equation} \label{eq:A1}
 d\xi= \bar n_{N}(x) dx, 
\end{equation}
where $\bar n_{N} \equiv n_{N}(x) / n_{N1}$ is the numerical density of neutrals normalised to its initial value.  Substituting $\xi$ for $n_N(x)$, Eqs.(\ref{eq:flux_mass})-(\ref{eq:flux_en}) (or Eqs.(\ref{eq:flux0_n})-(\ref{eq:flux0_mom}) for the relativistic case) can be rewritten by replacing $\partial_{x} \rightarrow \partial_{\xi}$, and using 
\begin{equation} \label{eq:A2}
  \dot n_1= n_{N1} n_{ph} \bar\sigma_{\rm ph} c  
\end{equation}
rather than $\dot n$ on the right-hand side.
This new set of equations is formally identical to Eqs.(\ref{eq:flux_mass})-(\ref{eq:flux_en}), where a constant density is assumed, with the only exception that $x$ is replaced by $\xi$. The solutions presented in \S\ref{sec:non-rel-model} and \S\ref{sec:rel-model} therefore remain the same, with the exception that the dependence on $x$ is substituted by the dependence on $\xi$.  To find the full new solutions therefore only requires that one finds the function $\xi(x)$, relating the new variable $\xi$ to the physical distance $x$. This can be easily done using Eq.(\ref{eq:A1}) along with the equation for the evolution of the neutral density, which, for the scenario A, is:
\begin{equation} \label{eq:A3}
 V_0 \frac{d n_N}{dx}= - \dot n = -n_N  n_{\rm ph} \bar\sigma_{\rm ph} c \,,
\end{equation}
leading to a simple exponential solution
\begin{equation} \label{eq:A4}
 n_N(x)= n_{N1} e^{-x/\lambda_{\rm ph}} \,,
\end{equation}
where $\lambda_{\rm ph}$ is defined in Eq.(\ref{eq:L_ph_ion}). Finally, substituting Eq.(\ref{eq:A4}) into Eq.(\ref{eq:A1}), and integrating one obtains
\begin{equation} \label{eq:A5}
 \xi(x) = \lambda_{\rm ph} \left( 1- e^{-x/\lambda_{\rm ph}} \right)  \,.
\end{equation}
It is easy to check that when $x \ll \lambda_{\rm ph}$ one has $\xi=x$, and that the solutions presented in \S\ref{sec:non-rel-model} and \S\ref{sec:rel-model} are recovered.

\section{Formation of secondary shocks}
\label{app:B}
One of the main results of this study is that the mass loading of a pulsar wind leads to a sideways expansion of the tail.  As the contact discontinuity between the tail flow of a PWN and the ISM acts as impenetrable wall for the ISM flow, a secondary shock can be created.  As we demonstrate below, this secondary shock will appear when the contact discontinuity makes an angle with the flow that is larger than the Mach cone. 

As a model problem, consider a flow with velocity, $V$, and internal sound speed, $c_s$, bounded by a wall with the parabolic profile, $r_0 = x_0^2/L$. A shock will appear when the characteristics first intersect \citep{LLVI}. The characteristics originating at point $\{x_0, r_0\}$ are 
\begin{equation}\label{eq:B1}
\begin{split}
 r(t)&=r_0 + c_s t,   \\
 x(t)&=x_0 + (c_s+V) t.
\end{split}
\end{equation}  
Eliminating $t$ leads to
\begin{equation} \label{eq:B2}
  r =  x_0^2/L+ \frac{c_s}{c_s+ V} (x-x_0).
\end{equation}
The first intersection occurs where $\partial r /\partial x_0 =0$:
\begin{equation}\label{eq:B3}
\begin{split}
 x_0 & = \frac{1}{2 (1+M)} L,   \\
 M & = V/c_s.
\end{split}
\end{equation}
At this point the angle that the wall makes with the flow is 
\begin{equation} \label{eq:B4}
  \tan \theta = 1/(1+M).
\end{equation}
Thus, for a highly supersonic flow a shock forms when the angle of attack becomes larger than the Mach angle. 

\section{Hydrodynamical model for relativistic wind} 
\label{app:C}
To derive Eqs. (\ref{eq:flux0_n})-(\ref{eq:flux0_mom}) we start by writing a relativistic formulation of the mass-loading problem, and then we specialise our equations to describe a typical pulsar wind. It has been shown by \cite{Komissarov94} \citep[see also][]{Lyutikov03} that the covariant relativistic equation for a perfect fluid with the inclusion of mass loading can be written in the following form:
\begin{equation} 
 \frac{\partial T^{\nu \mu}}{\partial x^{\nu}} = q c \tau^\mu \,,
\label{eq:rel_cons}
\end{equation}
where 
\begin{equation} 
 T^{\nu \mu}= w u^{\nu} u^{\mu} + P g^{\nu \mu}
\end{equation}
is the energy momentum tensor, $w=\epsilon+P$ is the total enthalpy, and $u^{\mu}=\gamma_w (1,{\bf u}/c)$ is the four-velocity of the plasma, with $\gamma_w$ denoting the Lorentz factor of the wind. The quantity $q c \tau^\mu$ represents the mass loading term, where $\tau^{\mu}=\gamma_{0}(1,{\bf V_{0}}/c)$ is the four-velocity of the neutrals moving with a Lorentz factor $\gamma_0=(1-V_0/c)^{-1/2}$, and 
\begin{equation}  \label{eq:q_inj}
 q=\sum_i \dot n_i \gamma_i m_i
\end{equation}
is the mass injected per unit time per unit volume calculated in the frame moving with velocity ${\bf V_0}$, \emph{i.e.}, the rest frame of the neutrals. It is assumed that the incoming neutrals have the same temperature as the ISM ($\approx 10^4$ K), and they can thus be considered as being cold with respect to the wind in the nebula, \emph{i.e}, $\gamma_i=1$.
As the rate of injected electrons and protons is the same,  $\dot n_e=\dot n_p = \dot n$, the equations describing the evolution of the electron and the proton densities are similar, \emph{i.e.},
\begin{equation} 
 \frac{\partial}{\partial x^{\nu}} \left[ n_{e,p} c u^{\nu} \right]= \dot{n} \,,
\label{eq:rel_cons_n}
\end{equation}
where $n_{e,p}$ is the numerical density of electrons (protons), with an expression for $\dot{n}$ given by Eq.(\ref{eq:dot_n}). As was done for the non-relativistic case, we assume that both $n_{N}$ and  $n_{ph}$ are constant along $x$. A method to obtain the solution when the depletion of neutrals inside the wind is taken into account is described in Appendix \ref{app:A}. 

The next step is to rewrite the conservation equations (\ref{eq:rel_cons}) and (\ref{eq:rel_cons_n}) for a 1-D system, integrating over a small volume with a cross section $A$ and a length $dx$ \cite[see][]{Komissarov94}. Additionally assuming that the system is in a steady-state, the conservation equations for the particles' number, energy and momentum in the $x$ direction become:
\begin{align} 
 \partial_x \left[ n_{p,e} u A \right]                            & = \dot n A' \,,  \label{eq:fluxC_n} \\
 \partial_x \left[ w \gamma_w u A \right]                  & = q c^2 \gamma_0 A' \,,   \label{eq:fluxC_en} \\
 \partial_x \left[ w u^2 A \right] + c^2 A\partial_x P  & = q c^2 \gamma_0 A' V_0 \,.  \label{eq:fluxC_mom}
\end{align}
Note that the non-relativistic equations (\ref{eq:flux_mass})-(\ref{eq:flux_en}) can be recovered from  Eqs.(\ref{eq:fluxC_n})-(\ref{eq:fluxC_mom}) using the first order approximation for the Lorentz factors, \emph{i.e.}, $\gamma_w \approx 1 + u^2/c^2$ and $\gamma_0 \approx 1 + V_0^2/c^2$.

\bibliographystyle{aa}
\bibliography{References_Morlino_Lyutikov_Vorster_2014}

\end{document}